\documentclass[12pt]{article}
\usepackage{epsfig}
 1
\def\as{\alpha_{\rm S}}
\def\aem{\alpha_{\rm em}}
\def\aems{\alpha^2_{\rm em}}
\def\dd{\partial}

\def\pom{{I\!\!P}}
\def\alphapom{{\alpha_{\pom}}}
\def\regg{{I\!\!R}}
\def\alpharegg{{\alpha_{\regg}}}
\def\citenum#1{{\def\@cite##1##2{##1}\cite{#1}}}
\def\citea#1{\@cite{#1}{}}

\def\lsim{\;\raisebox{-.4ex}{\rlap{$\sim$}} \raisebox{.4ex}{$<$}\;}

\def\as{\alpha_{\rm S}}

\def\gevs{{\,\mbox{GeV}^2}}
\def\to{\rightarrow}

\def\d{\delta}
\def\D{\Delta}

\def\l{\lambda}

\def\tim{\widetilde M}

\def\({\left(}
\def\){\right)}

\def\citenum#1{{\def\@cite##1##2{##1}\cite{#1}}}
\def\citea#1{\@cite{#1}{}}

\def\l1vt{\vec{l_{1\perp}}}

\def\bt{b_{\perp}}

\def\bkt{{\mathbf k_t}}
\def\bkt2{{\mathbf k^2_t}}
\def\bt2{$b^2_t$}

\def\jol1{$J_0(\,l_{1\perp}\,r_{\perp}\,)$}

\def\citea#1{\@cite{#1}{}}

\def\ie{\hbox{\it i.e.\ }}        \def\etc{\hbox{\it etc.\ }}
        
\def\etal{\hbox{\it et al.\ }}

\def\gtap{\ \raisebox{-.4ex}{\rlap{$\sim$}} \raisebox{.4ex}{$>$}\ }

\def\beq{\begin{equation}}
\def\eeq{\end{equation}}
\def\bea{\begin{eqnarray}}
\def\eea{\end{eqnarray}}

\def\eq#1{{\mbox{Eq.\hspace{1mm}(\ref{#1})}}}
\def\eqs#1#2{{\mbox{Eqs.\hspace{1mm}(\ref{#1})--(\ref{#2})}}}
\def\fig#1{{\mbox{Fig.\hspace{1mm}\ref{#1}}}}
\def\figs#1#2{{Figs.~\ref{#1}--\ref{#2}}}

\def\scrbox#1{\mbox{\scriptsize #1}}

\def\npb#1#2#3{    {\it Nucl.\ Phys.\ }{\bf B#1} (19#2) #3}
\def\plb#1#2#3{    {\it Phys.\ Lett.\ }{\bf B#1} (19#2) #3}
\def\prd#1#2#3{    {\it Phys.\ Rev.\ }{\bf D#1} (19#2) #3}
\def\prep#1#2#3{   {\it Phys.\ Rep.\ }{\bf #1} (19#2) #3}
\def\prl#1#2#3{    {\it Phys.\ Rev.\ Lett.\ }{\bf #1} (19#2) #3}

\def\zpc#1#2#3{    {\it Z.\ Phys.\ }{\bf C#1} (19#2) #3}

\def\yf#1#2#3{     {\it Yad.\ Fiz.\ }{\bf #1} (19#2) #3}
\def\sjnp#1#2#3{   {\it Sov.\ J.\ Nucl.\ Phys.\ }{\bf #1} (19#2) #3}
\def\jetp#1#2#3{   {\it Sov.\ Phys.\ }{JETP }{\bf #1} (19#2) #3}

\def\epj#1#2#3{    {\it Eur.\ Phys.\ J.\ }{\bf #1} (19#2) #3}

\def\ppjetp#1#2#3{ {\it (Sov.\ Phys.\ JETP }{\bf #1} (19#2) #3}

\newcommand{\email}[1]{${\!}^{\scrbox{#1)}}$}
\newcommand{\bm}[1]{\mbox{\boldmath{$#1$}}}
\newcommand{\lamlam}{{\lambda\lambda'}}
\def\bkj{\bm{k}_j}
\def\blj{\bm{\ell}_j}
\newcommand{\sig}[2]{\sigma^{\scrbox{#1}}_{\scrbox{#2}}}
\newcommand{\zf}[1]{\sum Z_{\scrbox{f}}^{#1}}
\newcommand{\qbar}{\overline{Q}^{\:2}}
\newcommand{\phit}{\varphi^{\scrbox{T}}}
\newcommand{\phil}{\varphi^{\scrbox{L}}}
\newcommand{\amp}{{\cal M}_{\lambda_1^{}\lambda_1'\lambda_2^{}\lambda_2'}}
\newcommand{\lratio}{\frac{\ell^2_1}{\ell^2_2}}
\newcommand{\ztola}[1]{\sqrt{1-4\frac{\ell_{#1}^2}{\tim_{#1}^2}}}
\newcommand{\ztolb}[1]{1-2\frac{\ell_{#1}^2}{\tim_{#1}^2}}
\newcommand{\sigtl}{\sigma^{\scrbox{T}}_{\scrbox{L}}}
\newcommand{\sigtt}{\sigma^{\scrbox{T}}_{\scrbox{T}}}
\newcommand{\siglt}{\sigma^{\scrbox{L}}_{\scrbox{T}}}
\newcommand{\sigll}{\sigma^{\scrbox{L}}_{\scrbox{L}}}

\relax
\begin{document}

\begin{titlepage}
\noindent
\begin{flushright}
\parbox[t]{10em}{
TAUP 2618/00\\
\today}
\end{flushright}

\vspace{1cm}
\begin{center}
  {\Large \bf The Components of the 
    ${\mathbf \gamma^* \gamma^*}$ Cross Section}
  \\[4ex]

\begin{minipage}{0.6\textwidth}
{\large E. ~G O T S M A N \email{1}, \hfill
        E. ~L E V I N     \email{2}, \vspace{2ex}\\
        U. ~M A O R       \email{3}  \hfill and \hfill 
        E. ~N A F T A L I \email{4}}
\end{minipage}
\footnotetext{\email{1} Email: gotsman@post.tau.ac.il .}
\footnotetext{\email{2} Email: leving@post.tau.ac.il .}
\footnotetext{\email{3} Email: maor@post.tau.ac.il .}
\footnotetext{\email{4} Email: erann@post.tau.ac.il .}
\\[4.5ex]
{\it  School of Physics and Astronomy}\\
{\it  Raymond and Beverly Sackler Faculty of Exact Science}\\
{\it  Tel Aviv University, Tel Aviv, 69978, ISRAEL}\\[4.5ex]

\end{center}
~\,\,
\vspace{1cm}

{\samepage {\large \bf Abstract:}} We extend our previous treatment of
$\gamma^{*}$p cross section based on Gribov's hypothesis to the case of
photon-photon scattering. With the aid of two parameters, determined from
experimental data, we separate the interactions into two categories
corresponding to short ("soft") and long ("hard") distance processes.  The
photon-photon cross section, thus, receives contributions from three sectors,
soft-soft, hard-hard and hard-soft. The additive quark model is used to
describe the soft-soft sector, pQCD the hard-hard sector, while the hard-soft
sector is determined by relating it to the $\gamma^{*}$p system. We calculate
and display the behaviour of the total photon-photon cross section and it's
various components and polarizations for different values of energy and
virtuality of the two photons, and discuss the significance of our results.

\end{titlepage}

\section {\label{section1} Introduction}
Scattering in the high energy (low $x$) limit has been studied in
perturbative QCD (pQCD) over the past few years, mainly through the analysis
of deep inelastic (DIS) events of lepton-hadron and hadron-hadron collisions.
Such pQCD investigation requires some knowledge of the non perturbative
contribution which is introduced, through the initial input to the evolution
equations or put in explicitly.  In this paper we present a study of virtual
photon-photon scattering. Our investigation is based on our model for
$\gamma^*$p cross section \cite{GLMN2} , which provides the framework for the
present calculation. Our goal is two fold:
\begin{enumerate}
\item In any QCD process, finding the dynamics for intermediate distances is
  still an open problem, as it involves a transition between short distance
  (``hard'' - perturbative) and large distance (``soft'' - non perturbative)
  physics. In Ref.\ \cite{GLMN2} we have suggested a procedure, based on
  Gribov's general approach \cite{GRIBOV}, of how to accommodate both
  contributions in DIS calculations. Two photon physics is an obvious
  reaction where these ideas can be further studied and re-examined.
\item Virtual photon-photon scattering has been proposed
  \cite{BRODSKY,BOONEKAMP,BARTELS,DESYLC} as a laboratory to study the BFKL
  Pomeron \cite{BFKL}, as the total cross section of two highly virtual
  photons provides a probe of BFKL dynamics. Our study enables one to
  estimate the background to the proposed BFKL process. This background
  consists of two contributions: ($\bm{i}$) We give an explicit estimate of
  the soft component in $\gamma^*\gamma^*$ scattering. ($\bm{ii}$) Our pQCD
  estimate for the hard component is based on DGLAP \cite{DGLAP} and
  as such can be used to assess when the BFKL dynamics start to dominate.
\end{enumerate}

Impressive attempts have been made \cite{SCHULER,BGHP} to describe two
  photon physics within the framework of Vector Dominance Model (VDM) mainly
  as a soft interaction. However, one can consider a two photon
  interaction as an interesting tool for investigating the interplay between
  soft and hard physics \cite{MSV2Pom}. The photon can appear as an unresolved
  object or as a perturbative fluctuation into an interacting quark-antiquark
  system. A careful analysis of the various components of the total cross
  section will help us understand the interface of the short distance and
  large distance interaction.
  
In $e^+e^-$ colliders, the measurement of the $\gamma^*\gamma^*$ is carried
  out by double tagging the outgoing leptons close to the forward direction,
  as most of the initial energy is taken by the scattered electrons.  The
  double tagged cross section falls off with the increase of the photons
  virtualities due to the photon propagator. The experimental statistics are
  improved for single and no tag events where one of the colliding photons or
  both are quasi-real \cite{DATAQ0}.  There is, therefore, a theoretical
  interest and an experimental need to better understand and estimate the
  perturbative and non-perturbative contributions with realistic
  configurations of the two photon virtualities.

Our paper deals with photon-photon collisions in the high energy limit, which
  confines us to low $x$ values. A pQCD investigation of $e\gamma$ DIS is non
  trivial \cite{NISIUSref} due to the dual nature of the photon target
  (quasi-real or virtual) which can be perceived as either a hadron like
  partonic system or a point like object. The resulting difficulties in pQCD
  calculations of $F_2^\gamma$ in the small $x$ limit have been extensively
  discussed in the literature and several strategies have been devised to
  bypass these problems \cite{NISIUSref}. For the purpose of our analysis we
  follow the approach suggested by Gl\"uck and Reya \cite{GR} in which the
  pQCD calculations has no predictive power regarding
  the normalization of
  $F_2^\gamma$ but it retains, as for a proton target, the $\as$ dependence
  of the evolution equations.
  
The above philosophy is very appropriate for our program where we
  distinguish between the hard pQCD mode and the non-perturbative QCD (npQCD)
  soft mode of the gluon fields by introducing \cite{GLMN2,GLMepj98}, two
  separation parameters ($M^2_{0,T}$ and $M^2_{0,L}$) in which we match the
  long and short distance components of the transverse (T) and longitudinal
  (L) contributions to the total $\gamma^*\gamma^*$ cross-section.  Our
  ideology is close to the Semiclassical Gluon Field Approach developed in
  Ref.\ \cite{MSV2}. This approach allows one to find a relation between
  scattering amplitudes and the property of the QCD vacuum based on the Model
  of the Stochastic Vacuum (MSV) \cite{MSV1}. Whereas the MSV is guided by
  the assumption of a microscopic structure of the QCD vacuum, our model is
  phenomenologically oriented based on the Additive Quark Model
  (AQM)\cite{AQM}.  The MSV has been combined \cite{MSV2Pom} with the two
  Pomeron model \cite{2Pom}.  In the two Pomeron model the hard Pomeron is a
  fixed $J$-pole whose $Q^2$ dependence is determined by fitting to data. In a
  pQCD calculation of the hard Pomeron, one has an effective $J$-pole whose
  dependence on $x$ and $Q^2$ is determined by $xG(x,Q^2)$. A short review of
  the various approaches to $\gamma^*\gamma^*$ reactions at high energies,
  stressing the need for a simultaneous determination of both the soft and
  the hard contributing components, has just appeared \cite{DSR00}.

The plan of our paper is as follows: In section \ref{section2} we review
  the generalization of the ideas presented in Ref.\ \cite{GLMN2} and outline
  the expansion of this model for the $\gamma^*\gamma^*$ cross section.  In
  section \ref{section3} we derive the complete set of formulae for the total
  cross section components. We present the details of our numerical
  calculations in section \ref{section4} and compare our results with the
  high energy experimental data available to date.  Our conclusions are
  summarized in section \ref{section5}.

\section{\label{section2} Review of the Approach}
Our approach follows from the ideas presented in Refs.\ \cite{BK}. This was
first suggested in Ref.\ \cite{GLMepj98} and successfully applied in Ref.\ 
\cite{GLMN2}.

\begin{figure}[tbp]
\begin{center}
  \epsfig{file=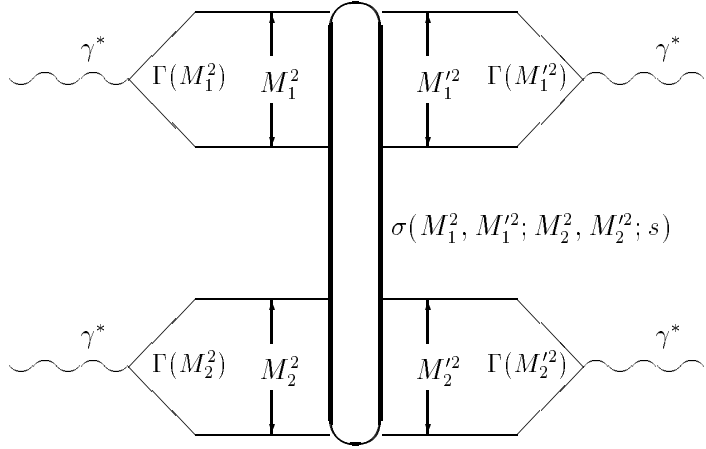,width=80mm,bbllx=200,bblly=550,bburx=450,bbury=700}
  \caption[]{\parbox[t]{
             0.80\textwidth}{\small \it Gribov's approach.}}
\label{fig1}
\end{center}
\end{figure}

According to Gribov's general approach \cite{GRIBOV}, the interaction of a
virtual photon, in any QCD description, can be interpreted as a two stage
process. The first stage is the fluctuation of the photon into a hadronic
system, and in the next stage the hadronic system interacts with the
``target'', which in our case is another hadronic system from a different
parent photon (see Fig.\ \ref{fig1}). These two processes are time ordered
and can be treated independently. The vertex function $\Gamma(M^2)$ of the
photon fluctuation into a $q\bar{q}$ pair of mass $M$ is given by the
experimental value of the ratio
\begin{equation}
\label{Rdef}
\Gamma(M^2) = R(M^2) = 
 \frac{\sigma(e^+e^-\to \mbox{hadrons})}{\sigma(e^+e^-\to \mu^+\mu^-)}\,.
\end{equation}

The complete process of two virtual photons fluctuating into two
quark-antiquark pairs which then interact with each other, can be expressed
by the following dispersion relation:
\begin{eqnarray}
\sigma(\gamma^*\gamma^*) &=& \left(\frac{\aem}{3\pi}\right)^2
\int dM^2_1\: dM^2_2\: dM'^2_1\: dM'^2_2 
\frac{\Gamma(M^2_1)}{\left(Q^2_1+M^2_1\right)}
\frac{\Gamma(M^2_2)}{\left(Q^2_2+M^2_2\right)} \times
\nonumber \\
& & \hspace{5cm}
\sigma(M^2_1,M'^2_1;M^2_2,M'^2_2;s)
\frac{\Gamma(M'^2_1)}{\left(Q^2_1+M'^2_1\right)}
\frac{\Gamma(M'^2_2)}{\left(Q^2_2+M'^2_2\right)}\,,
\label{disprel}
\end{eqnarray}
where $\sigma(M^2_1,M'^2_1;M^2_2,M'^2_2;s)$ is the cross section of the
interaction between two hadronic systems with masses $M_1$ and $M_2$ before
the interaction and $M'_1$ and $M'_2$ after the interaction.

We introduce a separation parameter in the mass integrations, which may be
different for longitudinal and transverse polarized virtual photon ($M_{0,L}$
and $M_{0,T}$ respectively). For masses below this parameter, the process is
soft, long range, and hence one cannot describe the produced hadron state as a
$q\bar{q}$ pair. For masses above the separation parameter, the distances
between the quark and antiquark are short, and
$\sigma(M^2_1,M'^2_1;M^2_2,M'^2_2;s)$, which depends on the gluon structure
function, can be calculated in pQCD.

\begin{figure}[tbp]
\begin{center}
  \epsfig{file=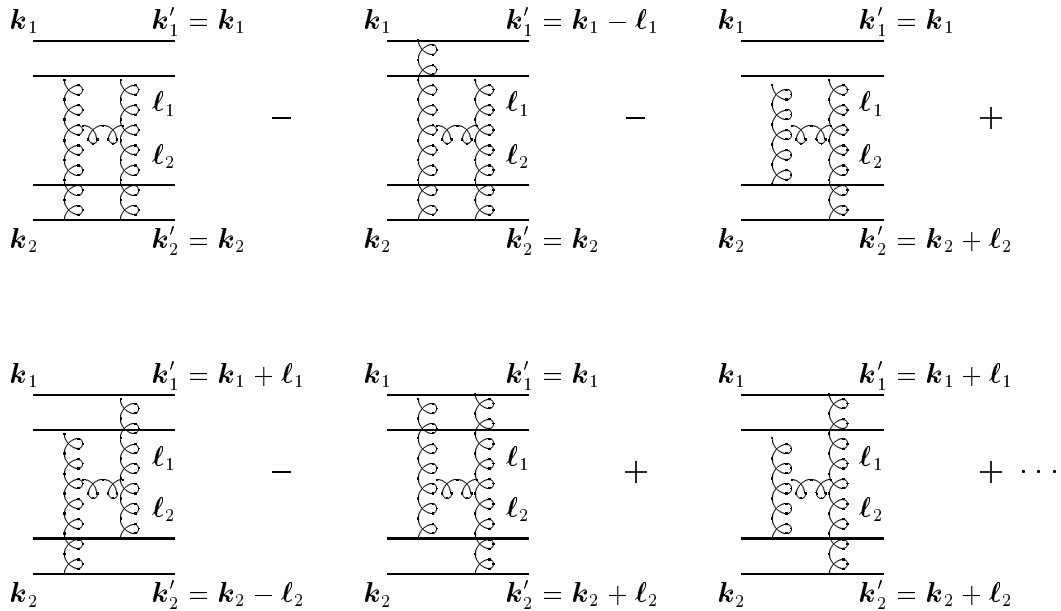,width=145mm,bbllx=100,bblly=500,bburx=560,bbury=780}
  \caption[]{\parbox[t]{
             0.80\textwidth}{\small \it Some of the diagrams which contribute to the pQCD
             calculations.}}
\label{fighh}
\end{center}
\end{figure}

The calculation of the two photon total cross section, according to our
approach, is derived following the same concepts of the $\sigma(\gamma^*p)$
calculations. Each of the two photons can be soft or hard, and we shall
derive the formulae on this basis. Without loss of generality, we shall
consider the case in which one photon (say, the upper one) is ``harder'' than
the other, hence there are three sectors of the calculation:
\begin{enumerate}
\item {\em ``Hard-hard''} when both photons are hard.  We treat the
  interaction between the two $q\bar{q}$ pairs in pQCD, calculating all the
  diagrams in which the upper $q\bar{q}$ pair are harder than the gluons in
  the ladder, and the gluons in the ladder are harder than the lower pair
  $k^2_1\gg\ell^2_1\gg\ell^2_2\gg k^2_2$ (see \fig{fighh}).  The cross
  section of the interaction in the hard-hard sector can be expressed through
  $xG_q$, the distribution function of a gluon ladder emitted from a single
  quark.

 To find $xG_q$ we recall that, in the region of small $x$, the
 evolution equation has the form:
 \begin{equation}
   \frac{d^2xG_q(x,Q^2)}{d\log\frac{1}{x}d\log Q^2} =
   \frac{N_c}{\pi}\,\alpha_S(Q^2)\,xG_q(x,Q^2).
 \end{equation}
 The solution\footnote{The function $xG_q$ has an additional term
   proportional to $\sqrt{\gamma(Q^2)(\log\frac{1}{x})^{-1}}\:
   K_{-1}\!\!\left(2\sqrt{\gamma(Q^2)\log\frac{1}{x}}\right)$, however at low
   $x$ this term contributes less than 1\% and can be neglected.} of this
 equation in the DLLA has been already given in
 \cite{ADL},
 \begin{equation}
    xG_q(x,Q^2) = G_0\,I_0\left(2\sqrt{\gamma(Q^2)\log\frac{1}{x}}\right),
    \label{xgq1}    
 \end{equation}
 with $G_0=0.0453$ and
 \begin{equation}
   \gamma(Q^2) =
   \frac{12N_c}{11N_c-2n_f}\log\left(
   \frac{\log\frac{Q^2}{\Lambda^2}}{\log\frac{Q^2_0}{\Lambda^2}}
   \right).
   \label{xgq2}
 \end{equation}
 
\item {\em ``Soft-soft''} for two soft photons. As stated, this is the case
  where neither of the hadronic systems can be treated in pQCD.  Here we use
  the AQM \cite{AQM} in which the interaction cross-section
  $\sigma(M^2_1,M'^2_1;M^2_2,M'^2_2;s)$ is diagonal with respect to $M_j$ and
  $M'_j\,(j=1,2)$.
  
\item {\em ``Hard-soft''} for the case that the upper photon is hard and the
  lower photon is soft. This sector is related, up to factorization, to the
  {\em hard} interaction between a nucleon and a photon, as the lower system
  is treated non perturbatively, while the upper hadronic system is a
  $q\bar{q}$ pair with small transverse separation. Thus, the
  interaction cross section $\sigma(M^2_1,M'^2_1;M^2_2,M'^2_2;s)$ is not
  diagonal and can be expressed through a nucleon gluonic structure function
  $xG$, with a factor of $\frac{2}{3}$, to account for the fact
  that we replace the nucleon $|qqq\rangle$ state with a $|q\bar{q}\rangle$
  state.
  
  As we shall see, our integrations require the knowledge of $xG(x,\ell^2)$
  where $\ell^2$ spans also over small values where the published
  parameterizations for $xG$ are not valid. We follow Refs.\ \cite{GLMN1}
  and \cite{GLMN2} and introduce an additional gluon scale $\mu^2$ and
  assume that the gluon structure function can be  approximated linearly by
  $\frac{\ell^2}{\mu^2}xG\left(x,\mu^2\right)$. Thus, our approach has two
  scales, one  separating the hard integration from the soft,
  which is related to the size of the quark-antiquark pair, and the gluon
  scale which is related to the size of the quark. For more details -- see
  Refs.\ \cite{GLMN2}, \cite{GLMN1}, and \cite{MRS98}.
 \end{enumerate}

In the next section, we derive explicitly the formulae for the three sectors
described above, taking into account both transverse and longitudinal
polarized photons in each sector. In our numerical calculations which are
presented in section \ref{section4}, the choice of our parameters are
consistent with Ref.\
\cite{GLMN2}. 

\section{\label{section3} Formulae for the Total $\bm{\gamma^*\gamma^*}$ 
Cross Section}

\subsection{The ``Hard-Hard'' components}

The pQCD calculation for the total cross section of two hard photons is
illustrated in diagrams of the type  shown in \fig{fighh}. We
denote
the production amplitude of the two systems of $q\bar{q}$, one from each
``hard'' photon, by $\amp$ where $\lambda_1\ldots\lambda_2'$ are the
helicities of the four quarks. We follow Refs.\cite{GLMN2,GLMepj98,LMRT} and
write $\amp$ in the form:
\begin{equation}
\amp = \sqrt{N_c}\int\; d^2\bm{k}_1'd^2z_1'\int\;d^2\bm{k}_2'd^2z_2'
                  {\cal T}_{1,2}\psi_1\psi_2
\end{equation}
where ${\cal T}_{1,2}={\cal T}(k'_1,k_1,k'_2,k_2)$ is the transition
amplitude of all the 16 possible diagrams of Fig.\ \ref{fighh},
$\psi_j=\psi_j(k_j',z_j')\,\, j=1, 2$ are the wave functions of the
$q\bar{q}$ inside the photons, and $z_j\;(1-z_j)$ is the fraction of the
energy of the $j$ photon that is carried by the quark(antiquark). Here and
throughout this paper our momentum variables are defined as the
two-dimensional transverse components of a four momentum, \ie $k\equiv
k_{\perp}$.

In the leading $\log(1/x)$ approximation $z=z'$ and therefore the transition
amplitude ${\cal T}_{1,2}$ does not dependent on $z_j,z_j'$. In the limit
where $k_1\gg\ell_1\gg\ell_2\gg k_2$ we write:
\begin{equation}
{\cal T}_{1,2} = i\frac{4\pi s}{2N_c}\int\frac{d^2\bm{\ell}_1}{\ell^2_1}
                                     \int\frac{d^2\bm{\ell}_2}{\ell^2_2}
                                     \D(k_1,k_1')\;\D(k_2,k_2')\;
                                     \as(\ell^2_1)\;f(x,\lratio)
\end{equation}
where,
\begin{equation}
\D(\bkj,\bkj')=2\d(\bkj'-\bkj)-\d(\bkj'-\bkj+\blj)-\d(\bkj'-\bkj-\blj)
\end{equation}
and $f$ is related to the gluons distribution function which will be
defined
below. Substituting ${\cal T}_{1,2}$ into $\amp$ and using the delta
functions, we get several combinations of products of the two wave functions,
\begin{equation}
\amp = i\frac{4\pi s}{2\sqrt{N_c}}\int\prod_j\frac{d^2\bm{\ell}_j}{\ell^2_j}
                                       \D\psi_{\lambda_j\lambda_j'}(k_j,z_j)\;
                                       \as(\ell^2_1)\; f(x,\lratio)
\label{amp}
\end{equation}
with
\begin{equation}
\Delta\psi(\bkj,z_j) = 2\psi(\bkj,z_j)-\psi(\bkj-\blj,z_j)-\psi(\bkj + \blj,z_j)\, .
\end{equation}
For the wave function of the $q\bar{q}$ pair inside a transverse and
longitudinally polarized photon, we shall use the results from Ref.\ 
\cite{LMRT}:
\begin{eqnarray}
\psi^{\pm}_\lamlam\left(\bkj ,z_j\right) 
   &=& -\delta_{\lambda , -\lambda'} Z_f e
                       \left[(1-2z_j)\lambda\mp 1\right] 
                 \frac{2\bm{\epsilon}_{\pm}\cdot\bkj}{\qbar_j+\bkj^2}
                          \hspace{1cm}\mbox{(transverse)}\, , \label{psiT}\\
\psi^{\scrbox{L}}_\lamlam(\bkj,z_j) &=& -2\delta_{\lambda , -\lambda'} Z_f e Q_j z_j(1-z_j)
                 \frac{1}{\qbar_j+\bkj^2} 
  \hspace{1cm}\mbox{ (longitudinal)}. \label{psiL} 
\end{eqnarray}
In (\ref{psiT}) and (\ref{psiL}), $Z_f$ is the charge of the quark
with flavour $f$ in units of the electron charge $-e$, $\qbar\equiv
z(1-z)Q^2$ and $\bm{\epsilon}_{\pm}=(0,0,1,\pm i)/\sqrt{2}$ is the photon
polarization vector. 

Carrying out the angular integration of $\Delta\psi$, we define the functions
$\varphi^{\scrbox{T}}$ and $\varphi^{\scrbox{L}}$ as follows:
\begin{eqnarray}
\lefteqn{
\int\,d^2\bm{\ell}
 \left[\frac{2\bm{\epsilon}_{\pm}\cdot\bm{k}}{\qbar+k^2}-
       \frac{\bm{\epsilon}_{\pm}\cdot(\bm{k}-\bm{\ell})}
         {\qbar+(\bm{k}-\bm{\ell})^2}-
       \frac{\bm{\epsilon}_{\pm}\cdot(\bm{k}+\bm{\ell})}
         {\qbar+(\bm{k+\ell})^2}
 \right]} & & \nonumber\\
& & =\,\pi\bm{\epsilon}_{\pm}\cdot\bm{k}\int\,d\ell^2\left[
\frac{\qbar-k^2}{\qbar+k^2} + 
\frac{\qbar-k^2+\ell^2}{\sqrt{(\qbar+k^2+\ell^2)^2-4k^2\ell^2}}\right] 
\nonumber\\
& & \equiv\, \pi\bm{\epsilon}_{\pm}\cdot\bm{k}
     \int\,d\ell^2\,\varphi^{\scrbox{T}}(k^2,\ell^2,\qbar)
\label{phiT}\\
& & \rule{0cm}{2ex} \nonumber \\
\lefteqn{
\int\,d^2\bm{\ell}
 \left[\frac{2}{\qbar+k^2}-
       \frac{1}{\qbar+(\bm{k}-\bm{\ell})^2}-
       \frac{1}{\qbar+(\bm{k}+\bm{\ell})^2}
 \right]} & & \nonumber\\
& & =\, 2\pi\int\,d\ell^2\left[
\frac{1}{\qbar+k^2}-
 \frac{1}{\sqrt{(\qbar+k^2+\ell^2)^2-4k^2\ell^2}}\right]
\nonumber\\
& & \equiv\, 2\pi\int\,d\ell^2 \varphi^{\scrbox{L}}(k^2,\ell^2,\qbar)
\label{phiL}
\end{eqnarray}
There are four hard-hard components for the two photon cross section, which
we  denote by $\sig{h(T)}{h(T)}$, $\sig{h(L)}{h(T)}$, \etc . 

We begin with the calculation of $\sig{h(T)}{h(T)}$.
Using the transition amplitude (\ref{amp}), the wave function (\ref{psiT})
and the angular integration (\ref{phiT}), we write:
\begin{eqnarray}
\lefteqn{
\sig{h(T)}{h(T)} = \frac{\zf{4}\aems}{\pi^4N_c}
\int\,dz_1\left[z^2_1+(1-z_1)^2\right]\int\,dz_2\left[z^2_2+(1-z_2)^2\right]}
\nonumber\\
& & \int\,\frac{dk^2_1}{\qbar_1+k_1^2}\,\int\,\frac{dk^2_2}{\qbar_2+k_2^2}
\int\,\frac{d\ell^2_1}{\ell^2_1}\,\int\,\frac{d\ell^2_2}{\ell^2_2}\,
\phit(k^2_1,\ell^2_1,\qbar_1)\,\phit(k^2_2,\ell^2_2,\qbar_2)\,
\as(\ell^2_1)f\left(x,\frac{\ell^2_1}{\ell^2_2}\right).
\label{htht1}
\end{eqnarray}

In order to perform the integration over
$z_1$ and $z_2$, we introduce the variables $M$ and $\tim$:
\begin{eqnarray}
M_j^2 &=& \frac{k_j^2}{z_j(1-z_j)} \nonumber \\
\tim_j^2 &=& \frac{\ell_j^2}{z_j(1-z_j)} \label{mmtdef}\, .
\end{eqnarray}
Formally, \eq{htht1} now has the form:
\begin{eqnarray}
\sig{h(T)}{h(T)} &=& \frac{\zf{4}\aems}{\pi^4N_c}\times
\nonumber\\
& &
\int\frac{dM^2_1}{Q^2_1+M_1^2}\,\int\frac{dM^2_2}{Q^2_2+M_2^2}
\int\frac{d\tim^2_1}{\tim^2_1}\,\int\frac{d\tim^2_2}{\tim^2_2}
\int\frac{d\ell^2_1}{\ell^2_1}\frac{\ztola{1}}{\ztolb{1}}\,
\int\frac{d\ell^2_2}{\ell^2_2}\frac{\ztola{2}}{\ztolb{2}}\times
\nonumber\\
& & 
\hspace{4.5cm}\phit(M^2_1,\tim^2_1,Q^2_1)\;\phit(M^2_2,\tim^2_2,Q^2_2)\;
\as(\ell^2_1)\;f\!\left(x,\frac{\ell^2_1}{\ell^2_2}\right).
\end{eqnarray}
 We now make some approximations:
\begin{enumerate}
\item In the limit $\ell_2^2 \gg k_2^2$ diagrams with
  $\bm{k_2}\ne\bm{k_2'}$ are suppressed, therefore we can neglect the
  integration over $\tim^2_2$ .
\item We can also simplify the $\ell_1^2$ and $\ell_2^2$ integration in the
  limits $\ell_1^2\ll\tim_1^2$ and $\ell_2^2\ll\tim_2^2$. The integrals are
  dominated by the upper integration limits dictated by the Jacobian, and we
  can safely replace $\ell^2_1$ and $\ell^2_2$ in $\as$ and $f$ by
  $\tim_1^2/4$ and $\tim_2^2/4$ respectively.
\item Integrating by parts over $\ell^2_1$, we redefine the gluon 
  ladder emitted by a quark:
  \begin{equation}
  \as(\ell_1^2)\,xG_q\!\left(x,\frac{\ell_1^2}{\ell^2_2}\right) = 
  \int^{\ell_1^2}\!\as(\ell_1^2)\,d\ell_1^2\,
  \!\int^{\ell^2_2}\frac{d\ell_2^2}{\ell_2^2}
   \,f\!\left(x,\frac{\ell_1^2}{\ell_2^2}\right).
   \label{xgqdef}
   \end{equation}
\end{enumerate}

Performing these simplifications we obtain:
\begin{eqnarray}
\sig{h(T)}{h(T)} &=& \frac{4\zf{4}\aems}{\pi^4N_c}
\int\,\frac{dM^2_1}{Q^2_1+M^2_1}\,\int\,\frac{dM^2_2}{Q^2_2+M_2^2}
\int\,\frac{d\tim^2_1}{\tim^4_1}\,\times
\nonumber\\
& & 
\hspace{5cm}
\as\left(\frac{\tim_1^2}{4}\right)\,xG_q\left(x,\frac{\tim_1^2}{M_2^2}\right)
\phit(M_1^2,\tim_1^2,Q_1^2).
\end{eqnarray}

As a last step, according to our approach, we set the limits of
the ``hard'' mass integrations, and replace each $2\zf{2}$ with the ratio
$R(M^2)$.  
\begin{eqnarray}
\sig{h(T)}{h(T)} &=& \frac{\aem^2}{\pi^4 N_c} 
              \int_{M_0^2}\frac{dM_1^2}{Q_1^2+M_1^2}\,R(M_1^2) 
              \int_{4m^2_\pi}{\frac{d\tim_1^2}{\tim_1^4}} 
              \as (\frac{\tim_1^2}{4})\, \phit(M_1,\tim_1,Q_1)\,\times
                                                            \nonumber\\
              & & \hspace{7cm}
              \int_{M^2_0}^{\tim^2_1}\frac{dM_2^2}{Q_2^2+M_2^2}\,R(M_2^2)\,
              xG_q(x,\frac{\tim_1^2}{M_2^2}). \label{htht}
\end{eqnarray}

The calculation of $\sig{h(L)}{h(T)}$ is straightforward. Using the same
assumptions, we find:
\begin{eqnarray}
\sig{h(L)}{h(T)} &=&  \frac{\aem^2}{4\pi^4 N_c} \,Q^2_1\,
              \int_{M_0^2}\frac{dM_1^2}{Q_1^2+M_1^2}\,R(M_1^2) 
              \int_{4m^2_\pi}{\frac{d\tim_1^2}{\tim_1^4}} 
              \as (\frac{\tim_1^2}{4})\, 
                      \varphi^L(M_1,\tim_1,Q_1)\,\times
                                            \nonumber\\
              & & \hspace{7cm}
              \int_{M^2_0}^{\tim^2}\frac{dM_2^2}{Q_2^2+M_2^2}\,R(M_2^2)\,
              xG_q(x,\frac{\tim_1^2}{M_2^2}). \label{hlht} 
\end{eqnarray}

We start our calculation of $\sig{h(T)}{h(L)}$, in the same way as we did for
the case of $\sig{h(T)}{h(T)}$, by collecting the transition amplitude
(\ref{amp}), the wave function (\ref{psiT}) and the angular integration
(\ref{phiT}):
\begin{eqnarray}
\lefteqn{
\sig{h(T)}{h(L)} = \frac{2\zf{4}\aems}{\pi^4N_c}\,Q_2^2
\int\,dz_1\left[z^2_1+(1-z_1)^2\right]\int\,dz_2\left[z_2\,(1-z_2)\right]^2}
\nonumber\\
& & \int\frac{dk^2_1}{\qbar_1+k_1^2}\,\int\frac{dk^2_2}{\qbar_2+k_2^2}
\int\frac{d\ell^2_1}{\ell^2_1}\,\int\frac{d\ell^2_2}{\ell^2_2}\,
\phit(k^2_1,\ell^2_1,\qbar_1)\;\phil(k^2_2,\ell^2_2,\qbar_2)\,
\as(\ell^2_1)f\left(x,\frac{\ell^2_1}{\ell^2_2}\right).
\label{hthl1}
\end{eqnarray}
We consider the limit $\ell^2_2\gg k^2_2\gg Q^2_2$ where
$\,\phil(k^2,\ell^2,\qbar) \longrightarrow \frac{1}{k^2_2}$. Using the
variables defined in \eq{mmtdef}, we find the lower photon part of
(\ref{hthl1}) to be:  
\begin{eqnarray}   
\lefteqn{Q^2_2\int dz_2\left\{\cdots\right\}\,
\int dk_2^2\left\{\cdots\right\}\,
\int d\ell_2^2\left\{\cdots\right\} =} & & \nonumber\\
& & \hspace{3cm}\int\frac{d\tim^2_2}{\tim^2_2\;(\ztolb{2})}\int\frac{dM^2_2}{Q^2_2+M^2_2}
\int\frac{d\ell^2_2}{\ell^2}\;\frac{Q^2_2\,\ell^2_2}{M_2^2\tim_2^2}.
\end{eqnarray}
The maximal value of $\ell^2_2$ is $\tim^2_2/4$ and the $\ell_2^2$
integral is dominated by that value. The lower photon part can now be written
in the form:
\begin{equation}
\int\frac{d\tim^2_2}{\tim^4_2}\int\frac{dM^2_2}{Q^2_2+M^2_2}
\;\frac{Q^2_2}{M_2^2}\label{hthl2}.
\end{equation}
Substituting (\ref{hthl2}) in (\ref{hthl1}), and switching the integration
variables of the upper photon $z_1,k^2_1$ into $M^2_1,\tim^2_1$ we have:
\begin{eqnarray}
\sig{h(T)}{h(L)} &\propto& 
\int\frac{dM^2_1}{Q^2_1+M^2_1}\int\frac{d\tim^2_1}{\tim^2_1}
\int\frac{d\ell^2_1}{\ell^2_1}\;\frac{\ztola{1}}{\ztolb{1}}\,
\as(\ell^2_1)\,\phit(M^2_1,\tim^2_1,Q^2_1) \times
\nonumber\\
& &
\hspace{5cm}
Q^2_2\,\int\frac{dM^2_2}{M^2_2(Q^2_2+M^2_2)}\int\frac{d\tim^2_2}{\tim^4_2}\,
f\!\left(x,\frac{4\ell^2_1}{\tim^2_2}\right).
\end{eqnarray}
We can now use the definition (\ref{xgqdef}) of 
$xG_q$, and perform an integration by parts over $\tim^2_2$:
\begin{equation}
\int_{M^2_2}^{\tim^2_1}\frac{d\tim^2_2}{\tim^4_2}\, f = 
\frac{1}{M^2_2}\,xG_q - \int_{M^2_2}^{\tim^2_1}\frac{d\tim^2_2}{\tim^4_2}xG_q\,.
\label{mt2int}
\end{equation}
Finally we integrate by parts over $\ell^2_1$ and obtain, in the limit
$\tim^2_1 \gg 4\ell^2_1$:
\begin{eqnarray}
\sig{h(T)}{h(L)} &=&  \frac{2\aem^2}{\pi^4 N_c}\,Q^2_2\,
              \int_{M_0^2}\frac{dM_1^2}{Q_1^2+M_1^2}\,R(M_1^2) 
              \int_{4m^2_\pi}{\frac{d\tim_1^2}{\tim_1^4}} 
              \as (\frac{\tim_1^2}{4})\, \varphi^T(M_1,\tim_1,Q_1)\,\times
                                                            \nonumber\\
              & & \hspace{0cm}
              \int_{M^2_0}^{\tim^2_1}\frac{dM_2^2}{M^2_2(Q_2^2+M_2^2)}\,
                                                                   R(M_2^2)\,
              \left\{\frac{1}{M^2_2}xG_q(x,\frac{\tim_1^2}{M_2^2}) \,-\, 
                     \int_{M^2_2}^{\tim^2_1}\frac{d\tim_2^2}{\tim_2^4}
                             xG_q(x,\frac{\tim_1^2}{\tim_2^2})\right\} \,.
                                                                \label{hthl}
\end{eqnarray}
Following the same procedure, we obtain the last term of the ``hard-hard''
sector, 
\begin{eqnarray}
\sig{h(L)}{h(L)} &=&  \frac{\aem^2}{2\pi^4 N_c}\,Q^2_1 Q^2_2\,
              \int_{M_0^2}\frac{dM_1^2}{Q_1^2+M_1^2}\,R(M_1^2) 
              \int_{4m^2_\pi}{\frac{d\tim_1^2}{\tim_1^4}} 
              \as (\frac{\tim_1^2}{4})\, \varphi^L(M_1,\tim_1,Q_1)\,\times
                                                            \nonumber\\
              & & \hspace{0cm}
              \int_{M^2_0}^{\tim^2_1}\frac{dM_2^2}{M^2_2(Q_2^2+M_2^2)}\,
                                                                   R(M_2^2)\,
              \left\{\frac{1}{M^2_2}xG_q(x,\frac{\tim_1^2}{M_2^2}) \,-\, 
                     \int_{M^2_2}^{\tim^2_1}\frac{d\tim_2^2}{\tim_2^4}
                             xG_q(x,\frac{\tim_1^2}{\tim_2^2})\right\} .
              \label{hlhl} 
\end{eqnarray}

\subsection{The ``Soft-Soft'' Components}
When the $q\bar{q}$ pair for both photons have small invariant masses, the
distance between the quark and antiquark is long, and following \cite{GLMN2},
we use the AQM and write the cross section for the interaction
of one soft hadron state with another soft hadron state, as
\begin{equation}
\sigma\left(M^2_1,M'^2_1;M^2_2,M'^2_2;s\right) = 
\sigma^{\scrbox{soft}}\left(M^2_1,M^2_2,s\right)
  \delta\left(M^2_1-M'^2_1\right)\delta\left(M^2_1-M'^2_1\right).
\end{equation}
In \cite{GLMN2} we dealt with a photon proton interaction and used the
pion-proton cross section for parameterizing $\sigma^{\scrbox{soft}}$. The
soft cross section between two photons is related to $\sigma(\pi p)$ by a
factor of $\frac{2}{3}$, which comes from the fact that there are {\em two}
$q\bar{q}$ systems as opposed to the the $\pi p$ case where one $q\bar{q}$
pair interacts with a nucleon. This approach is valid in the region of small
$x < 0.1$. For the $\pi p$ cross section we  use the Donnachie -
Landshoff Reggeon parameterization \cite{DL},
\begin{equation}
\sigma_{DL} = A_{\pom} \left(\frac{M_1^2\,M_2^2}{s\,s_0}\right)^{-\alphapom+1} + 
              A_{\regg} \left(\frac{M_1^2\, M_2^2}{s\, s_0}\right)^{-\alpharegg+1}.
\end{equation}

As in the case of hard-hard contributions, the {\em total} soft-soft cross
section also has four terms, one for each of the four possible polarizations
of the two photons. These terms will be denoted by, $\sig{s(T)}{s(T)},\,
\sig{s(T)}{s(L)}\,$ \etc  The contributions from each soft photon are
($j=1,2$): 
\begin{equation}
\int_{4m_{\pi}^2}^{M^2_{0,T}}\frac{R(M_j^2)M_j^2dM_j^2}{(Q_j^2+M_j^2)^2}
\hspace{1em}\mbox{for transverse photon,}
\label{softT}
\end{equation}
and,
\begin{equation}
\int_{4m_{\pi}^2}^{M^2_{0,L}} \frac{R(M_j^2)Q_j^2dM_j^2}{(Q_j^2+M_j^2)^2} \hspace{1em}\mbox{for longitudinal photon.}
\label{softL}
\end{equation}
Note that the limits of integration differ for the different polarizations
(see \cite{GLMN2}). Below we summarize the formulae for the soft-soft
components:
\begin{eqnarray}
\sigma^{s(T)}_{s(T)} &=& \left(\frac{\aem}{3\pi}\right)^2 
                       \int_{4m_{\pi}^2}^{M^2_{0,T}}
                        \frac{M_1^2\,R(M_1^2) dM_1^2}{\left(Q_1^2+M_1^2\right)^2}
                       \int_{4m_{\pi}^2}^{M^2_{0,T}}
                        \frac{M_2^2\,R(M_2^2) dM_2^2}{\left(Q_2^2+M_2^2\right)^2}
                       \,\,\frac{2}{3}
                       \sigma_{DL}\, ;
                        \\
                       & & \rule{0cm}{2ex} \label{stst} \nonumber \\
\sigma^{s(L)}_{s(T)} &=& \left(\frac{\aem}{3\pi}\right)^2 
                       \int_{4m_{\pi}^2}^{M^2_{0,L}}
                        \frac{Q_1^2\,R(M_1^2) dM_1^2}{\left(Q_1^2+M_1^2\right)^2}
                       \int_{4m_{\pi}^2}^{M^2_{0,T}}
                        \frac{M_2^2\,R(M_2^2) dM_2^2}{\left(Q_2^2+M_2^2\right)^2}
                       \,\,\frac{2}{3}
                       \sigma_{DL}\, ;
                        \\
                       & & \rule{0cm}{2ex} \label{slst} \nonumber \\
\sigma^{s(T)}_{s(L)} &=& \left(\frac{\aem}{3\pi}\right)^2 
                       \int_{4m_{\pi}^2}^{M^2_{0,T}}
                        \frac{M_1^2\,R(M_1^2) dM_1^2}{\left(Q_1^2+M_1^2\right)^2}
                       \int_{4m_{\pi}^2}^{M^2_{0,L}}
                        \frac{Q_2^2\,R(M_2^2) dM_2^2}{\left(Q_2^2+M_2^2\right)^2}
                       \,\,\frac{2}{3}
                       \sigma_{DL}\, ;
                        \\
                       & & \rule{0cm}{2ex} \label{stsl} \nonumber \\
\sigma^{s(L)}_{s(L)} &=& \left(\frac{\aem}{3\pi}\right)^2 
                       \int_{4m_{\pi}^2}^{M^2_{0,L}}
                        \frac{Q_1^2\,R(M_1^2) dM_1^2}{\left(Q_1^2+M_1^2\right)^2}
                       \int_{4m_{\pi}^2}^{M^2_{0,L}}
                        \frac{Q_2^2\,R(M_2^2) dM_2^2}{\left(Q_2^2+M_2^2\right)^2}
                       \,\,\frac{2}{3}
                       \sigma_{DL}\,  .
                        \\
                       & & \rule{0cm}{1ex} \label{slsl} \nonumber
\end{eqnarray}

\subsection{The ``Hard-Soft'' components}
We now consider the case for the interaction of a hard photon (the
upper) and a soft photon (the lower). We shall take into account both
polarizations of the soft photon together, and split it into the two terms
(\ref{softT}) and (\ref{softL}) at the end of this section. We start with the
case in which the hard photon is transverse polarized, and denote this term
temporarily as $\sig{h(T)}{s}$. A priori, our expression for
$\sig{h(T)}{s}$ can
be written in the form:
\begin{eqnarray}
\sig{h(T)}{s} &=&
\frac{2\aems\zf{2}}{3\pi^2N_c}\int dz_1\left[z^2_1+(1-z_1)^2\right]\times
\nonumber\\
& &
                              \int\frac{dk^2_1}{\qbar_1+k^2_1}
                              \int\frac{d\ell^2_1}{\ell^4}\phit(k^2_1,\ell^2_1,\qbar_1)\;
                                  \as(\ell^2_1)f\left(x,\frac{\ell^2_1}{Q^2_0}\right)
                              \int_{\scrbox{soft}}\frac{R(M^2_2)dM^2_2}{Q^2_2+M^2_2}.
\label{hts1}
\end{eqnarray}
We see that in \eq{hts1} the two photon cross section factorizes, and we
shall deal, for the moment, with the hard piece alone,
\begin{equation}
\zf{2}\int\frac{dM^2_1}{Q^2_1+M^2_1}
      \int\frac{d\tim^2_1}{\tim^2_1}
      \int\frac{d\ell^2_1}{\ell^4_1}
        \frac{\ztola{1}}{\ztolb{1}}\phit(M^2_1,\tim^2_1,Q^2_1)
        \as(\ell^2_1)f\left(x,\frac{\ell^2_1}{Q^2_0}\right)
\end{equation}
where (\ref{mmtdef}) had been used to change variables from $z_1,\,
k^2_1,\, \ell^2_1$ into $M^2_1,\, \tim^2_1,\, \ell^2_1$. In the limit
$4\ell^2_1\ll\tim^2_1$ we can integrate by parts over $\ell^2_1$ and obtain:
\begin{equation}
\frac{1}{2}\int_{M_{0,T}^2}\frac{R(M^2_1)dM^2_1}{Q^2_1+M^2_1}
           \int 4\frac{d\tim_1^2}{\tim_2^4}\as\left(\frac{\tim_1^2}{4}\right)
                \frac{2}{3}xG\left(x,\frac{\tim_1^2}{4}\right) .
\label{hts2}
\end{equation}
In (\ref{hts2}) we replaced $\ell_1^2$ in the argument of $\as$ and $f$ with
the dominant point of the integration range and used the following equation
as the definition of the gluon distribution function:
\begin{equation}
\int^{\ell^2}\frac{d\ell^2}{\ell^2}\as(\ell^2)f\left(x,\frac{\ell^2}{Q_0^2}\right)
= \frac{2}{3}xG(x,\ell^2)\, ,
\label{xgdef}
\end{equation}
where $xG(x,\ell^2)$ is the gluon distribution function in a {\em nucleon} and
it can be taken from one of the existing parameterizations \cite{GRV98,MRS99}.

Substituting (\ref{hts2}) in (\ref{hts1}) we
write the expression for the interaction of hard transverse photon with soft
photon in the form:
\begin{eqnarray}
\sigma^{h(T)}_s &=&  \frac{4\aem^2}{3\pi^2 N_c} 
              \int_{\scrbox{soft}}
                \frac{dM_2^2}{Q_2^2+M_2^2}\,R(M_2^2) 
              \int_{M^2_{0,T}}\frac{dM_1^2}{Q_1^2+M_1^2}\,R(M_1^2) 
              \int_{4m^2_\pi}{\frac{d\tim_1^2}{\tim_1^4}} \,\times
              \nonumber\\
              & & \hspace{6cm}
              \as (\frac{\tim_1^2}{4})\; \frac{2}{3}\,xG(x,\frac{\tim_1^2}{4})
               \,\varphi^T(M_1,\tim_1,Q_1)\,. \label{hts3}
\end{eqnarray}
Notice that the limits of soft integration are specified in \eqs{softT}{softL}. As
stated, we shall separate the soft piece at the end of the section, but until
then we denote it as ``$\int_{\scrbox{soft}}$''.

The last case is the one where the hard photon is longitudinally
polarized. The formal expression for the cross section is similar to
\eq{hts1},
\begin{eqnarray}
\sig{h(L)}{s} &=&
\frac{4\aems\zf{2}}{3\pi^2N_c}\int dz_1\left[z_1(1-z_1)\right]^2\times
\nonumber\\
& &
                              Q^2_1\int\frac{dk^2_1}{\qbar_1+k^2_1}
                              \int\frac{d\ell^2_1}{\ell^4}\phil(k^2_1,\ell^2_1,\qbar_1)\;
                                  \as(\ell^2_1)f\left(x,\frac{\ell^2_1}{Q^2_0}\right)
                              \int_{\scrbox{soft}}\frac{R(M^2_2)dM^2_2}{Q^2_2+M^2_2}.
\label{hls1}
\end{eqnarray}
Using (\ref{mmtdef}) to change the integration variables into 
$M^2_1,\,\tim^2_1$ and $z_1$, we write the hard piece of (\ref{hls1}) in the
form:
\begin{equation}
\zf{2}\int dz_1
      \int\frac{dM^2_1}{Q^2_1+M^2_1}
      \int\frac{d\tim^2_1}{\tim^4_1}\;\phit(M^2_1,\tim^2_1,Q^2_1)\;
           \as\left(z_1(1-z_1)\tim^2_1\right)f\!\left(x,z_1(1-z_1)\tim^2_1\right)\,.
\label{hls2}
\end{equation}
Integrating by parts over $\tim$ we obtain, using (\ref{xgdef}):
\begin{eqnarray}
\sig{h(L)}{s} &=&  \frac{2\aem^2}{3\pi^2 N_c}\,Q^2_1\, 
              \int_{\scrbox{soft}}\frac{dM_2^2}{Q_2^2+M_2^2}\,R(M_2^2) 
              \int_{M^2_0}\frac{dM_1^2}{Q_1^2+M_1^2}\,R(M_1^2) 
              \int_{4m^2_\pi}{\frac{d\tim_1^2}{\tim_1^2}} \,\times
              \nonumber\\
              & & \hspace{5cm}
              \frac{2}{3}\,\overline{xG}(x,\tim_1^2)
              \left[\frac{1}{\tim_1^2} - \frac{\dd}{\dd\tim_1^2}\right]
              \varphi^L(M_1,\tim_1,Q_1)\, ,
              \label{hls3}
\end{eqnarray}
where,
\begin{equation}
\overline{xG}(x,\ell^2) =
\int^1_0\as\!\left(z(1-z)\ell^2\right) xG\!\left(x,z(1-z)\ell^2\right)\:dz\,.
\label{xgbardef}
\end{equation}

In  our approach, we use different cutoffs for transverse and
longitudinal photons, thus the soft pieces of Eqs.\ (\ref{hts3}) and
(\ref{hls3}) -- the lower photon --
are  separated into the two terms of \eqs{softT}{softL},
\begin{equation}
\int_{\scrbox{soft}}\frac{R(M^2_2)}{Q^2_2+M^2_2}dM^2_2\longrightarrow
\int_{4m_{\pi}^2}^{M^2_{0,T}}\frac{M_2^2\;R(M^2_2)}{(Q^2_2+M^2_2)^2}dM^2_2 +
\int_{4m_{\pi}^2}^{M^2_{0,L}}\frac{Q_2^2\;R(M^2_2)}{(Q^2_2+M^2_2)^2}dM^2_2\,.
\label{splitsoft}
\end{equation}
Substituting (\ref{splitsoft}) in $\sig{h(T)}{s}$ and $\sig{h(L)}{s}$, we
finally obtain the set of four formulae in the ``hard-soft'' sector:
\begin{eqnarray}
\sig{h(T)}{s(T)} &=&  \frac{4\aem^2}{3\pi^2 N_c} 
              \int_{4m^2_\pi}^{M^2_{0,T}}
                \frac{M_2^2\,dM_2^2}{\left(Q_2^2+M_2^2\right)^2}\,R(M_2^2) 
              \int_{M^2_{0,T}}\frac{dM_1^2}{Q_1^2+M_1^2}\,R(M_1^2) 
              \int_{4m^2_\pi}{\frac{d\tim_1^2}{\tim_1^4}} \,\times
              \nonumber\\
              & & \hspace{6cm}
              \as (\frac{\tim_1^2}{4})\; \frac{2}{3}\,xG(x,\frac{\tim_1^2}{4})
               \,\phit(M_1,\tim_1,Q_1) \, ;\label{htst} \\
           & & \rule{0cm}{2ex} \nonumber \\
\sig{h(T)}{s(L)} &=&  \frac{4\aem^2}{3\pi^2 N_c} 
              \int_{4m^2_\pi}^{M^2_{0,L}}
                \frac{Q_2^2\,dM_2^2}{\left(Q_2^2+M_2^2\right)^2}\,R(M_2^2) 
              \int_{M^2_{0,T}}\frac{dM_1^2}{Q_1^2+M_1^2}\,R(M_1^2) 
              \int_{4m^2_\pi}{\frac{d\tim_1^2}{\tim_1^4}} \,\times
              \nonumber\\
              & & \hspace{6cm}
              \as (\frac{\tim_1^2}{4})\; \frac{2}{3}\,xG(x,\frac{\tim_1^2}{4})
               \,\phit(M_1,\tim_1,Q_1) \, ;\label{htsl} \\
           & & \rule{0cm}{2ex} \nonumber \\
\sig{h(L)}{s(T)} &=&  \frac{2\aem^2}{3\pi^2 N_c}\,Q^2_1\, 
              \int_{4m^2_\pi}^{M^2_{0,T}}
              \frac{M_2^2\,dM_2^2}{\left(Q_2^2+M_2^2\right)^2}\,R(M_2^2) 
              \int_{M^2_{0,L}}\frac{dM_1^2}{Q_1^2+M_1^2}\,R(M_1^2) 
              \int_{4m^2_\pi}{\frac{d\tim_1^2}{\tim_1^2}} \,\times
              \nonumber\\
              & & \hspace{5.5cm}
              \frac{2}{3}\,\overline{xG}(x,\tim_1^2)
              \left[\frac{\dd}{\dd\tim_1^2} -
              \frac{1}{\tim_1^2}\right]\phil(M_1,\tim_1,Q_1)\, ,
              \label{hlst} \\
              & & \rule{0cm}{2ex} \nonumber \\
\sig{h(L)}{s(L)} &=&  \frac{2\aem^2}{3\pi^2 N_c}\,Q^2_1\, 
              \int_{4m^2_\pi}^{M^2_{0,L}}
              \frac{Q_2^2\,dM_2^2}{\left(Q_2^2+M_2^2\right)^2}\,R(M_2^2) 
              \int_{M^2_{0,L}}\frac{dM_1^2}{Q_1^2+M_1^2}\,R(M_1^2) 
              \int_{4m^2_\pi}{\frac{d\tim_1^2}{\tim_1^2}} \,\times
              \nonumber\\
              & & \hspace{5.5cm}
              \frac{2}{3}\,\overline{xG}(x,\tim_1^2)
              \left[ \frac{1}{\tim_1^2} -
              \frac{\dd}{\dd\tim_1^2}\right]\phil(M_1,\tim_1,Q_1)\, .
              \label{hlsl}
\end{eqnarray}

\section{\label{section4} Numerical Calculations}

Our final expression for the total cross section is a sum of all terms
  for the three sectors derived in section \ref{section3}:
  \begin{equation}
     \sigma(\gamma^*\gamma^*) = \sigma(\scrbox{Hard-Hard}) + 
                                \sigma(\scrbox{Hard-Soft}) + 
                                \sigma(\scrbox{Soft-Soft}).
     \label{total}\end{equation}
  Where each of the components in (\ref{total}) includes terms from all
  possible polarizations, with two contributions for each ``non-diagonal''
  term.  In all our numerical calculations we have used values for the
  parameters (within the error bounds) that are consistent with those that
  had been introduced for the proton photon cross section \cite{GLMN2} \ie
  $0.7<M^2_{0,T}<0.9 \gevs$, $M^2_{0.L}\lsim 0.4 \gevs$.  An additional
  parameter in the present calculation is $Q^2_0$ which appears in
  \eq{xgq2}, we set $Q^2_0=0.48 \gevs$.  In the integration over very low
  masses we assume \cite{GLMN2,GLMepj98} that the gluon structure functions,
  both in the hard-hard sector and in the hard-soft sector behave as
  $xG\left(x,\ell^2<\mu^2\right)=\frac{\ell^2}{\mu^2}xG\left(x,\mu^2\right)$,
  where $\mu^2=0.8 \gevs$ for $xG^{\scrbox{GRV}}$ and $\mu^2=Q^2_0$ for
  $xG_q$.
   
\begin{figure}
\begin{center}
  \epsfig{file=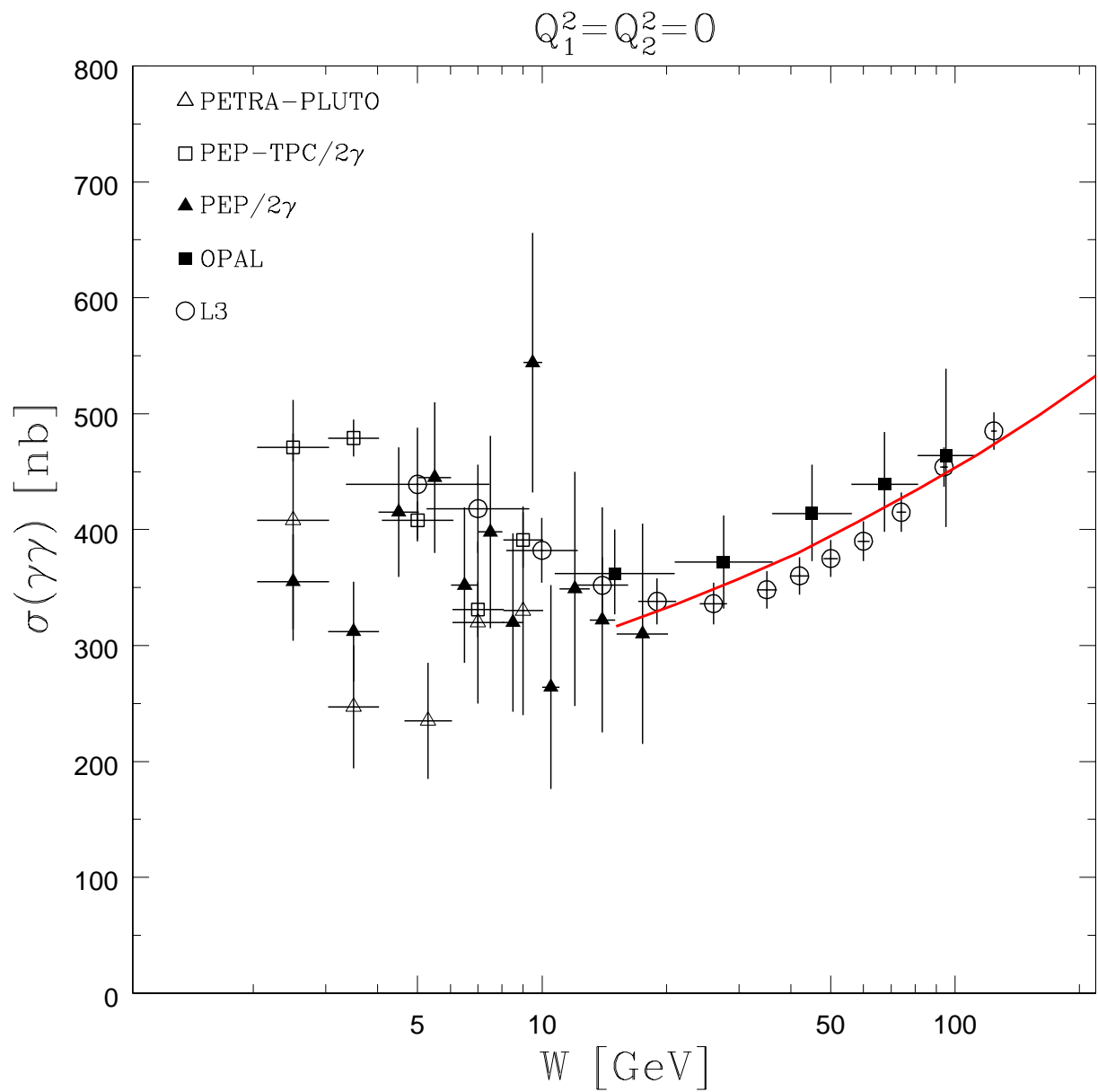,width=\textwidth}
  \caption[]{\parbox[t]{
             0.80\textwidth}{\small \it Our calculations for $\sigma(\gamma\gamma)$ and the
             experimental data.}}
\label{qeq0}
\end{center}
\end{figure}

We initially compare our numerical calculations with the published
   experimental data \cite{DATAQ0,OPAL,L3,L3DTAG}, in which one photon is
   always real or semi real (\ie $Q^2_2=0$). These are presented in
   \figs{qeq0}{q14}, where our calculation and experimental data are plotted
   as a function of $W$ for fixed $Q^2_1=0,\,3.5$ and $14 \gevs$. In the
   region of high energy, we are in good agreement with data at each of the
   measured virtualities. For real photons, the cross section is dominated
   by the soft-soft sector, while for the $\gamma^*\gamma$ case the hard-soft
   contribution increases, and at high energies ($\gtap 50\gevs$) become
   larger than the soft-soft component [see \fig{softhard}]. The hard-hard
   component is small at high energy even for high values $Q^2_1$. We are
   unable to reproduce the experimental low energy enhancement which is
   clearly observed in the data displayed in \figs{qeq0}{q14}. This is a
   direct consequence of the threshold enhancement associated with the point
   like photon component which is not included in our calculations.
   
The total cross section is dramatically reduced when the second photon is
   virtual. In \fig{Q2eq1} we show for comparison the results of the
   numerical calculation for two processes at fixed $Q^2_1$ and $Q^2_2$. 
   The
   first is the collision of a highly virtual photon ($Q^2_1=3.5,\, 14$ and
   $20\gevs$) with a real photon, and the second is the collision of the highly
   virtual with a semi-hard photon with $Q^2_2=1 \gevs$. When the second
   photon is semi-hard, the total cross section is smaller by a factor of 5-7
   for $Q^2_1=3.5 \gevs$ and by a factor of 6-8 for $Q^2_1=20 \gevs$. The
   contribution of the hard-hard sector cannot be neglected when
   $Q^2_2=1\gevs$, as can be seen in \fig{softhardq1}, where we show the
   percentage of the three sectors for the semi-hard second photon for fixed
   values of $Q^2_1$. This figure is to be compared with \fig{softhard}.

It is instructive to study the $Q^2$ dependence of our formulae as well. We
  fixed the value of $W$ and $Q^2_2$ and calculated the total cross section
  term by term. It figure \fig{fixw} we show the total cross section for two
  values of $Q^2_2$ and five values of $W$, and in \fig{shfixw} we plot the
  percentage of the three sectors at fix $Q^2_2$ and $W$. It is worth noting
  that at high energies and non-zero virtualities the hard-hard sector
  contribute up to 20\% to the total cross section.
  
As far as the photon's polarization is concerned, our formalism enables us
  to define the total cross section as a sum of four expressions:
  $\sigtt,\,\sigtl,\,\siglt$ and $\sigll$, where each of these components
  contain contributions from all the three sectors. For $Q^2_2=0$,
  $\sigtl=\sigll=0$ and we are left with only two non-vanishing components
  which we can use to define the ratio between longitudinal and transverse
  cross section as $\sigma^L/\sigma^T = \siglt/\sigtt$. This ratio is shown
  in \fig{ltreal}, as a function of the energy for constant values of
  $Q^2_1$, where the decreasing of the longitudinal component with the energy
  can be seen. Note that at high energies this ratio approaches the value of
  $\approx 0.35$ and it does not depend on the value of $Q^2_1$. However, if
  the second photon has  non-zero virtuality, all of the four components
  contribute to the total cross section. In \fig{ltvirt} we present the
  relative contribution of the polarization components as a percentage from
  $\sigma(\gamma^*\gamma^*)$ both for $Q^2_2=0$ and $Q^2_2=1 \gevs$. The
  contribution to the cross section from the component of two longitudinal
  photons is less than 5\%.

\section{\label{section5} Conclusions}

In this paper we have presented a detailed calculation of the total cross
section for the collision of two virtual photons. Our calculations are based
upon a generalization of Gribov's formula for $\gamma^{*}$p scattering which
has been amended for the two photon case.  In our approach the
parameterization of the two photon interaction is a natural extension of
$\gamma$ proton scattering.

We introduce two separation parameters $M^{2}_{0T}$ and $M^{2}_{0L}$
which allows us to stipulate the long and short distance components of the
interaction. With the aid of these parameters we are able to separate
the two photon cross section into three sectors: soft-soft, hard-soft and
hard-hard. This enables us to investigate the interplay between the large
and short distance processes, and to evaluate their relative contributions
to the total photon-photon cross section. Our calculation of the soft sector
is based on the DL parameterization. The hard sector is calculated in pQCD
utilizing DGLAP.

The main conclusions of this paper are:
\begin{enumerate}
\item For one or two quasi-real photons our model describes the data in the
  region of high energy (small $x$). Though the $Q^2_1=Q^2_2=0$ interaction
  of two photons is mainly soft, it receives a contribution from what we call
  the hard-soft sector, which is a signature of short distance processes for
  one of the photons. On the other hand, even if one photon is highly
  virtual, the soft-soft sector does not vanish, and therefore npQCD effects
  also contribute to the hard photon sector.
\item In the case where both photons are off shell, the total cross section
  is considerably smaller, and the contribution of the hard-hard sector
  cannot be neglected. This effect occurs already for intermediate distances
  of the order of $1\gevs$.
\item Comparing $\sigma(\gamma^*\gamma)$ with $\sigma(\gamma^*\gamma^*)$ we
  observe a somewhat steeper rise of the former as a function of $W$. This
  stems from the hard-hard sector, which has a moderate energy behaviour, and
  becomes important only for two off shell photons at high energies.
\item Each of the three sector is a sum of four components for all possible
  polarizations of the photons.  For a realistic situation in which one
  photon has low virtuality, the cross section is dominated by the
  transversely polarized photons. Terms in which both photons are
  longitudinally polarized are small ($< 5 \%$), while those for mixed
  polarization are not negligible.
\item The hopes that photon-photon physics in LEP2 and near future $e^+e^-$
  colliders will serve as a clear probe of BFKL dynamics depend on relatively
  small background from the soft sector and DGLAP hard sector.  Not
  withstanding the expected low statistics of double tagged experiments, we
  note that these contributions are rather significant at realistic values of
  the interacting photons virtualities even at high energies.
\end{enumerate}

{\large \bf Acknowledgments:} This research was supported in part by the
Israel Science Foundation, founded by the Israel Academy of Science and
Humanities.

\begin{figure}
\begin{center}
  \epsfig{file=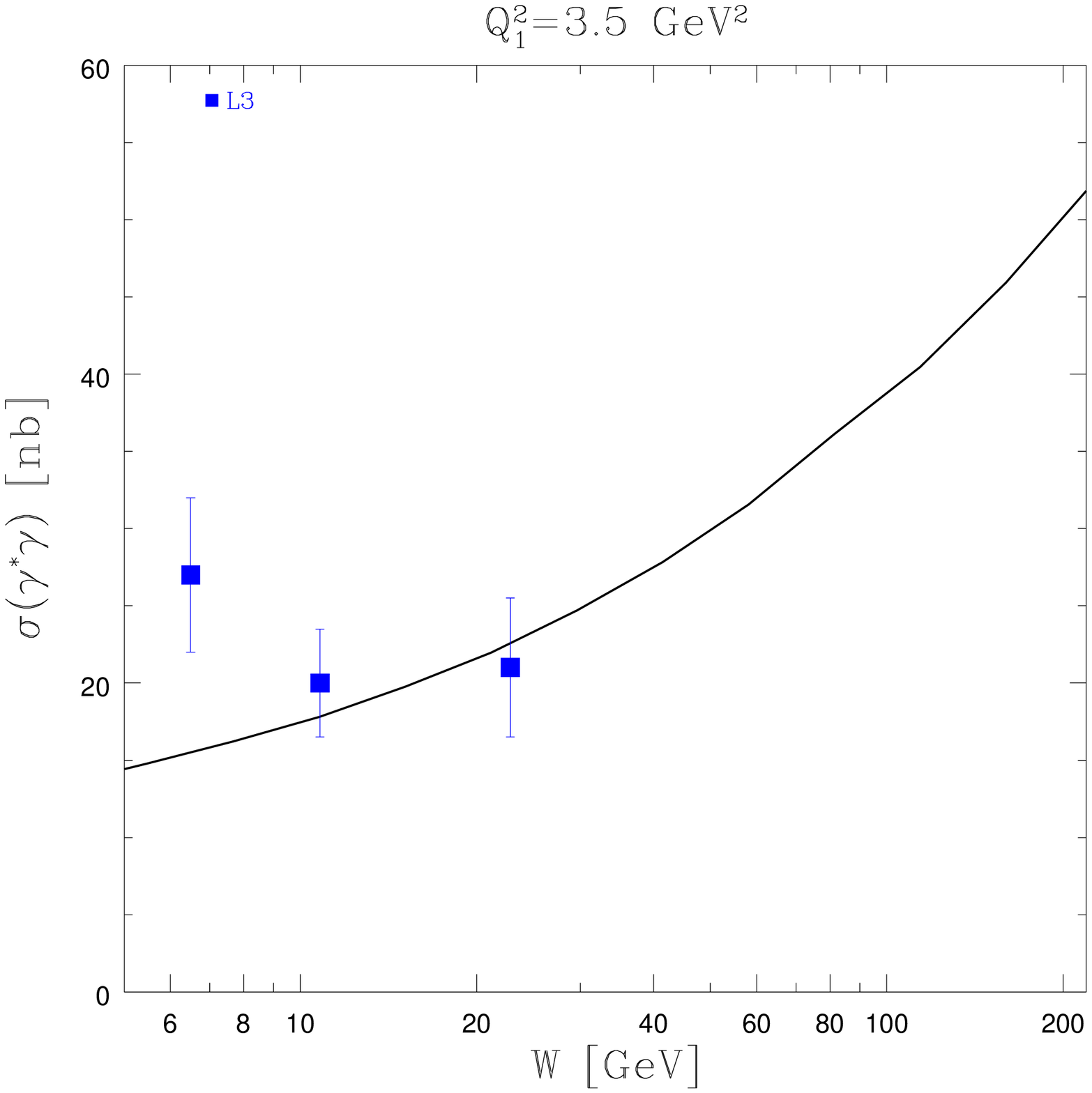,width=\textwidth}
  \caption[]{\parbox[t]{
             0.80\textwidth}{\small \it Our calculations for $\sigma(\gamma^*\gamma)$ and the
             experimental data for $Q^2_1=3.5 \gevs$ and $Q^2_2=0$.}}
\label{q35}
\end{center}
\end{figure}

\begin{figure}
\begin{center}
  \epsfig{file=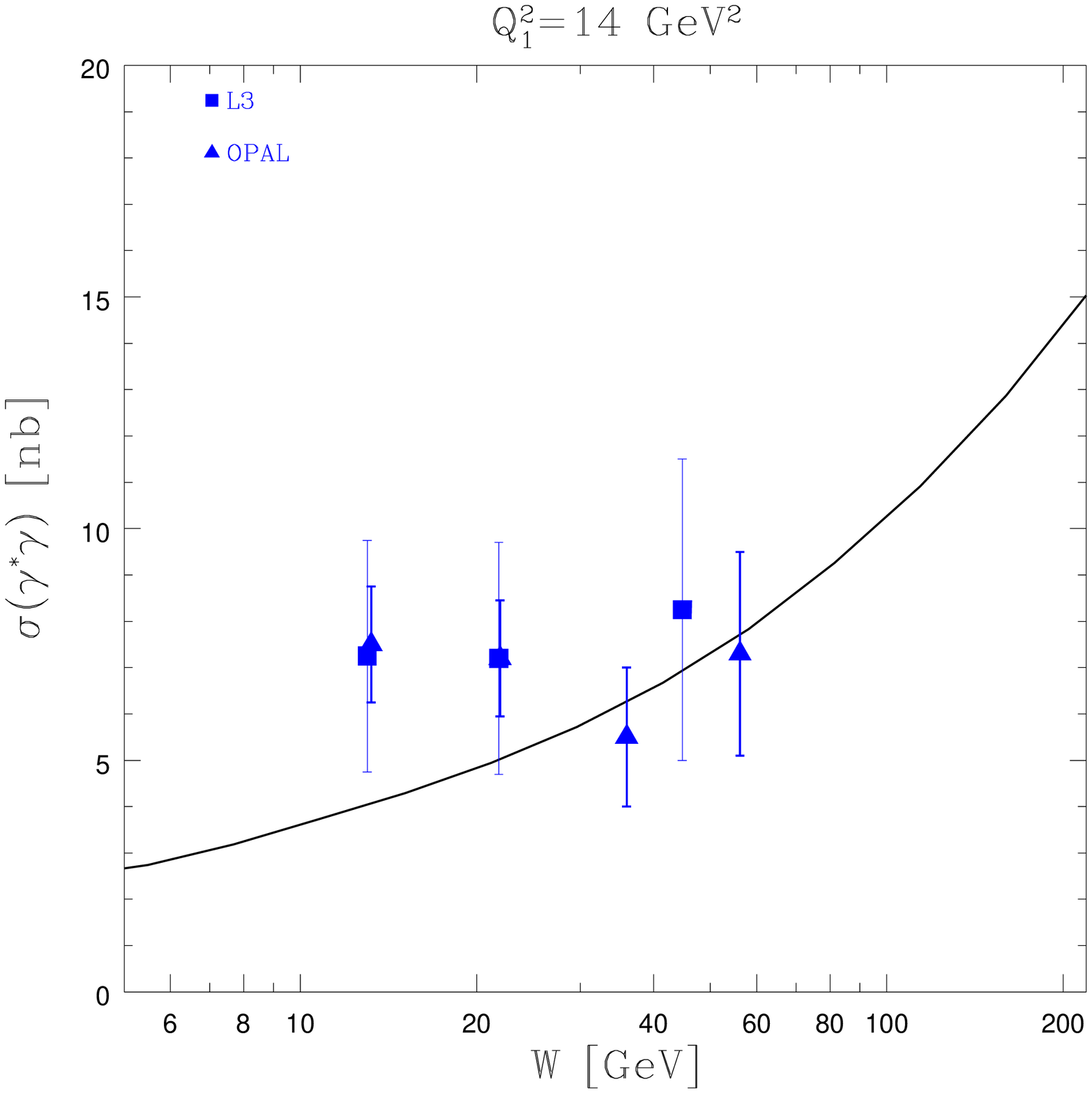,width=\textwidth}
  \caption[]{\parbox[t]{
             0.80\textwidth}{\small \it Our calculations for $\sigma(\gamma^*\gamma)$ and the
             experimental data for $Q^2_1=14 \gevs$ and $Q^2_2=0$.}}
\label{q14}
\end{center}
\end{figure}

\begin{figure}
\begin{center}
  \epsfig{file=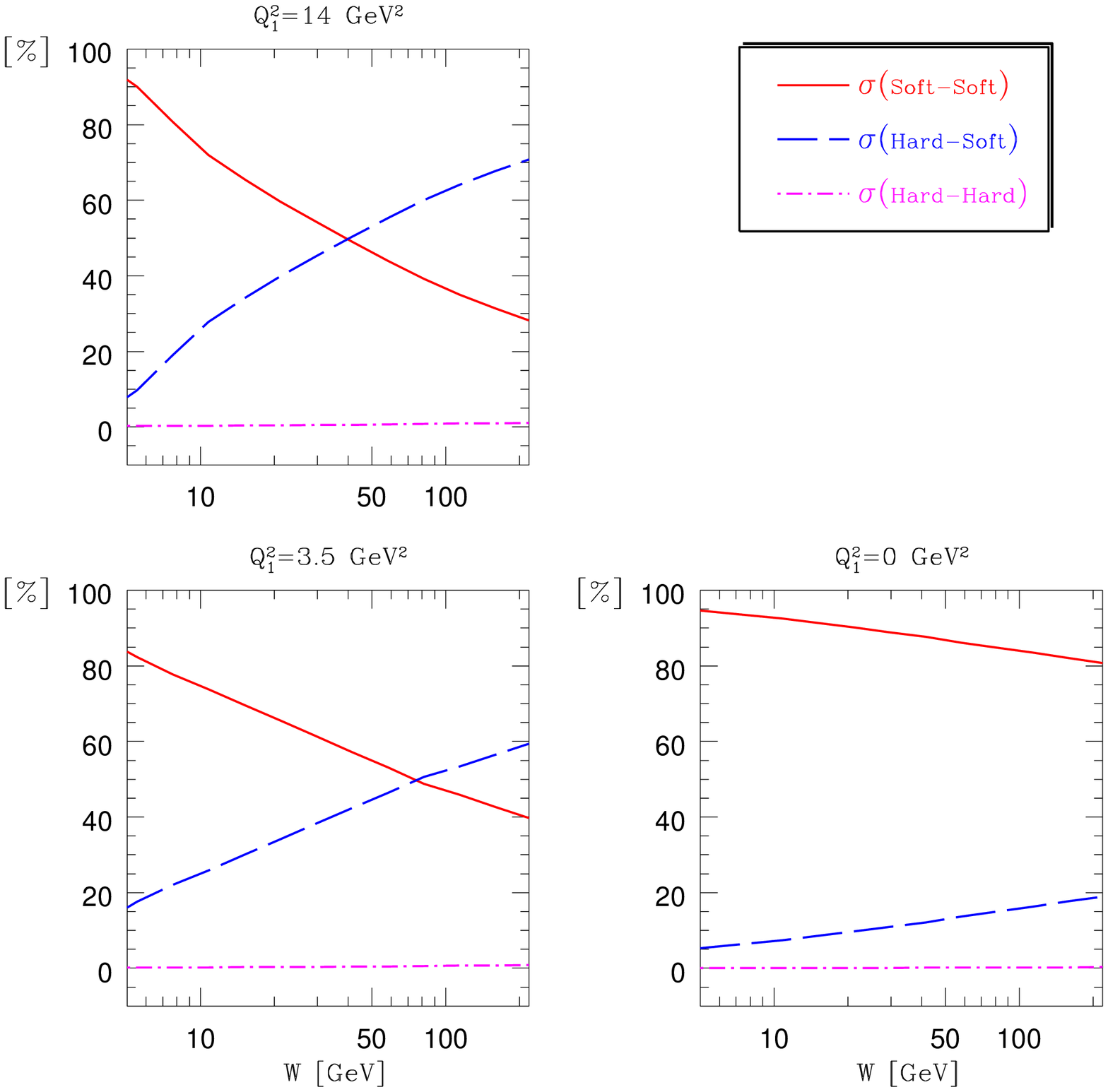,width=\textwidth}
  \caption[]{\parbox[t]{
             0.80\textwidth}{\small \it The relative contribution of
             the three sectors as a percentage from
             $\sigma(\gamma^*\gamma^*)$ as a function of $W$, for the case of
             $Q^2_2=0$.}}
\label{softhard}
\end{center}
\end{figure}

\begin{figure}
\begin{center}
  \epsfig{file=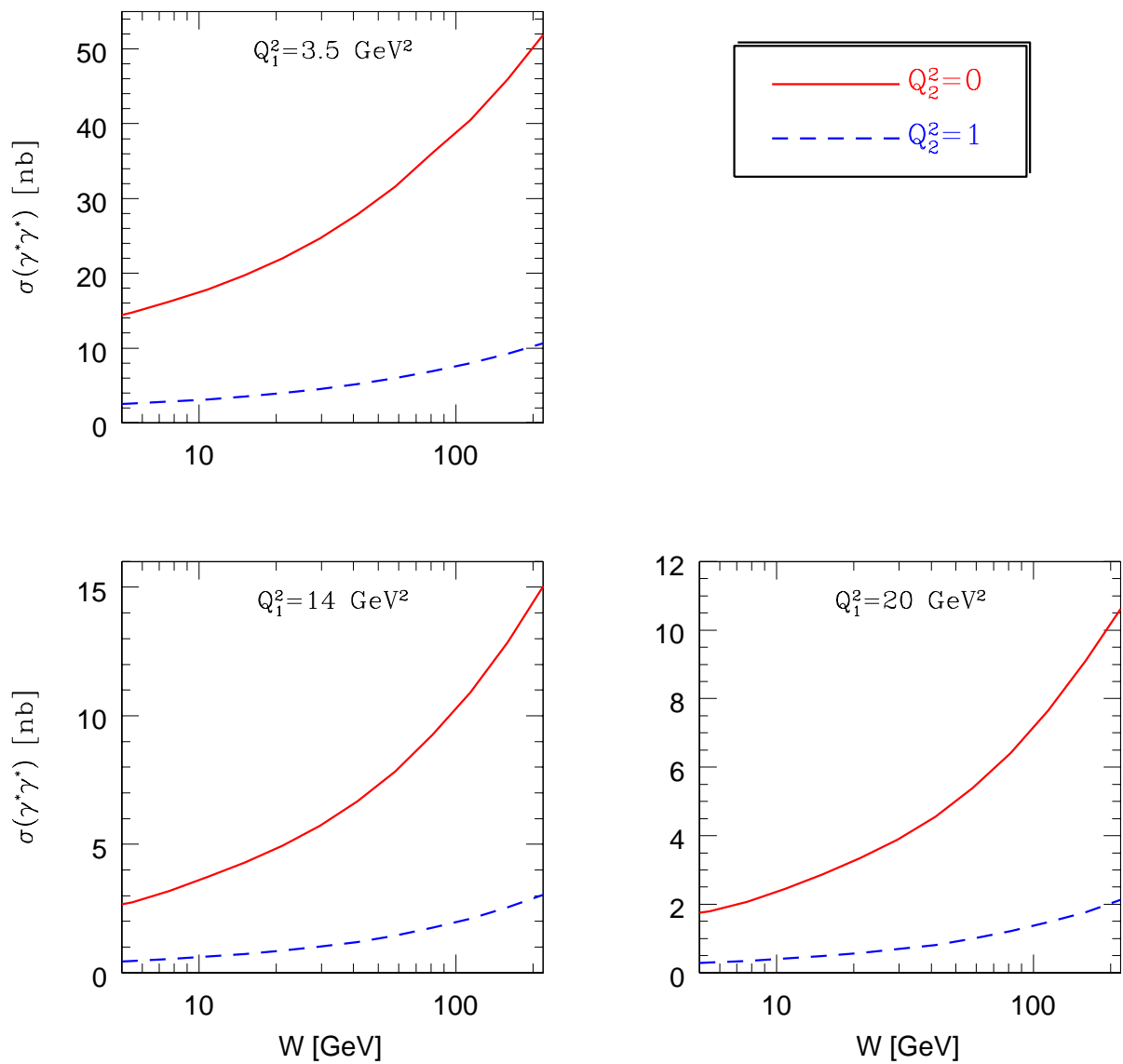,width=\textwidth}
  \caption[]{\parbox[t]{
             0.80\textwidth}{\small \it $\sigma(\gamma^*\gamma^*)$ for fixed
             values of the two photons virtualities.}}
\label{Q2eq1}
\end{center}
\end{figure}

\begin{figure}
\begin{center}
  \epsfig{file=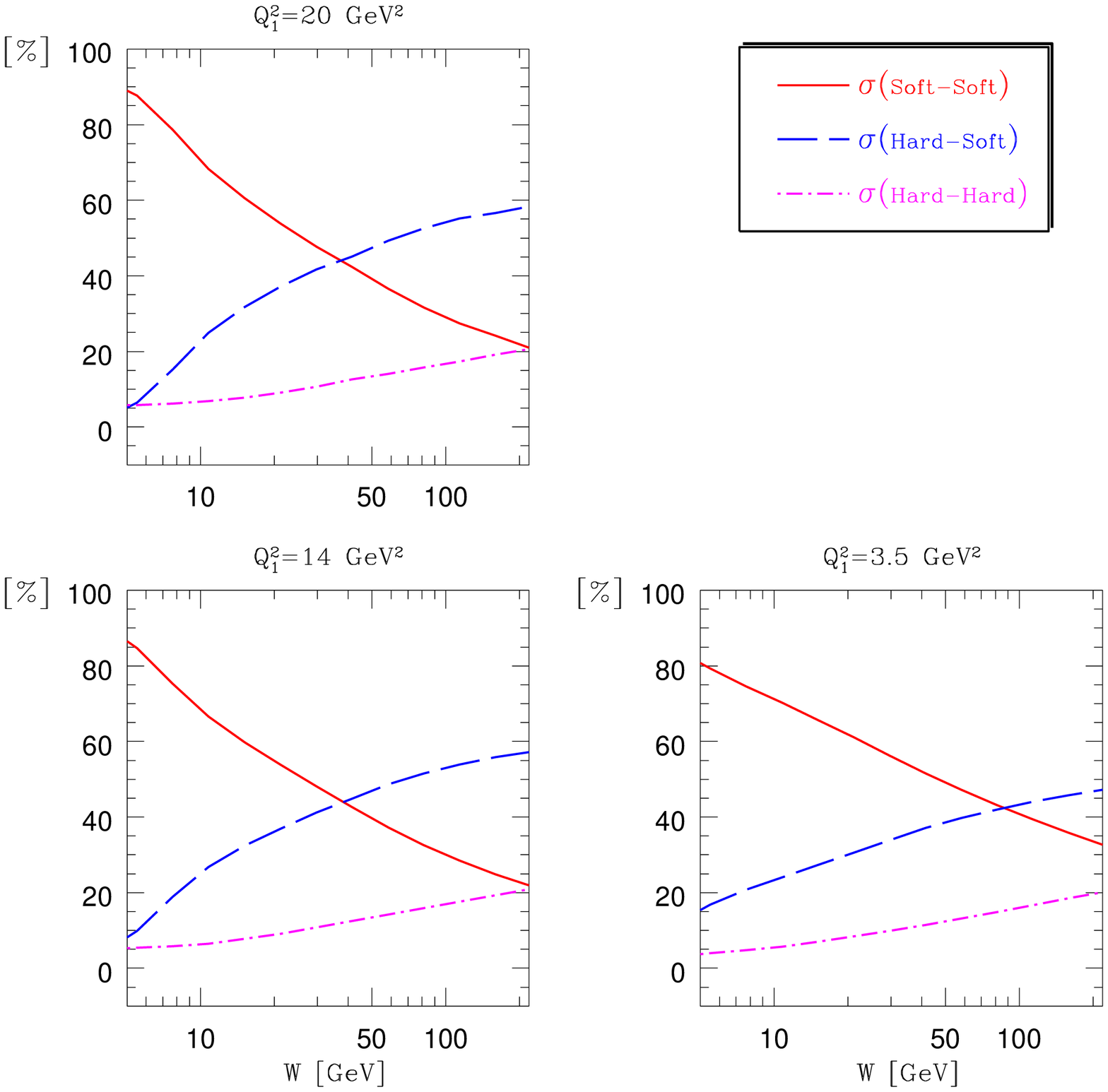,width=\textwidth}
  \caption[]{\parbox[t]{
             0.80\textwidth}{\small \it The relative contribution of
             the three sectors as a percentage from
             $\sigma(\gamma^*\gamma^*)$ as a function of $W$, for the case of
             $Q^2_2=1$.}}
\label{softhardq1}
\end{center}
\end{figure}

\begin{figure}
\begin{center}
  \epsfig{file=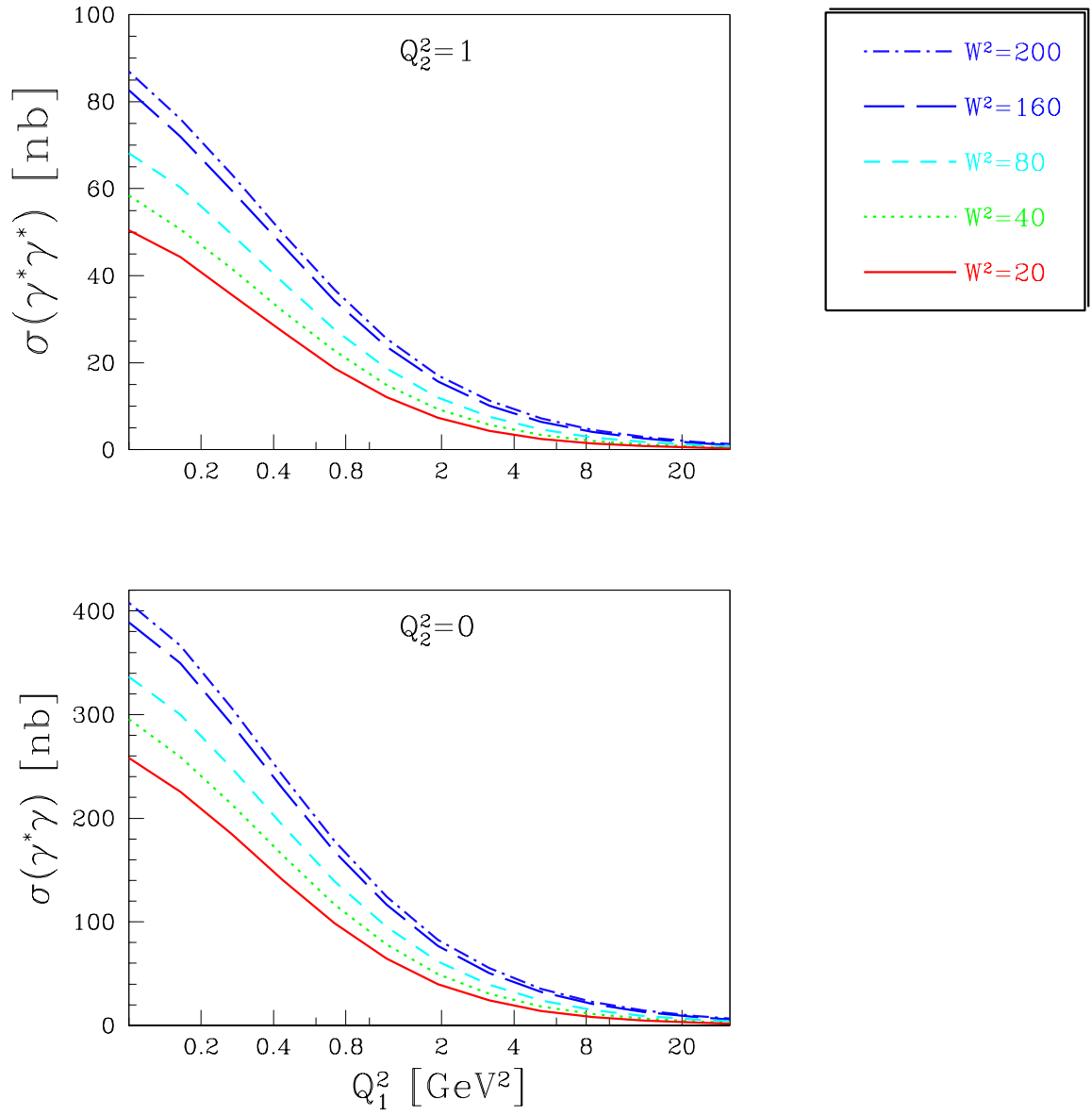,width=\textwidth}
  \caption[]{\parbox[t]{
             0.80\textwidth}{\small \it $\sigma(\gamma^*\gamma^*)$ as a
             function of $Q^2_1$ for fixed $W$ and two different values of $Q^2_2$.}}
\label{fixw}
\end{center}
\end{figure}

\begin{figure}
\begin{center}
  \epsfig{file=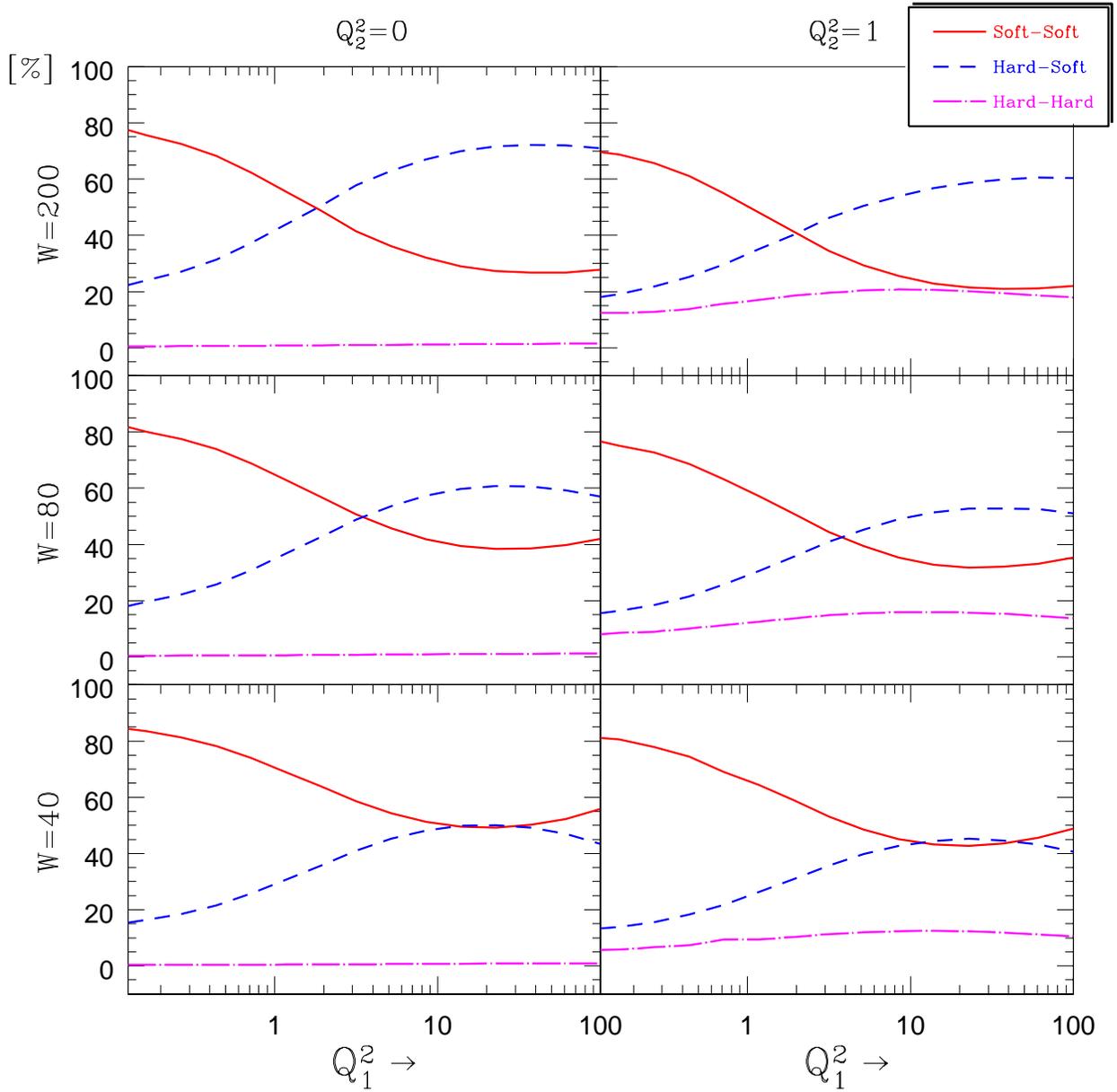,width=\textwidth}
  \caption[]{\parbox[t]{
             0.80\textwidth}{\small \it The relative contribution of
             the soft-soft, hard-soft and hard-hard as a percentage from
             $\sigma(\gamma^*\gamma^*)$ for fix $W$ and $Q^2_2$. The three
             graphs on the left column are for $Q^2_2=0$ and right column
             correspond to $Q^2_2=1$.
             The values of $W$ are shown on the left hand side of the figure.}}
\label{shfixw}
\end{center}
\end{figure}

\begin{figure}
\begin{center}
  \epsfig{file=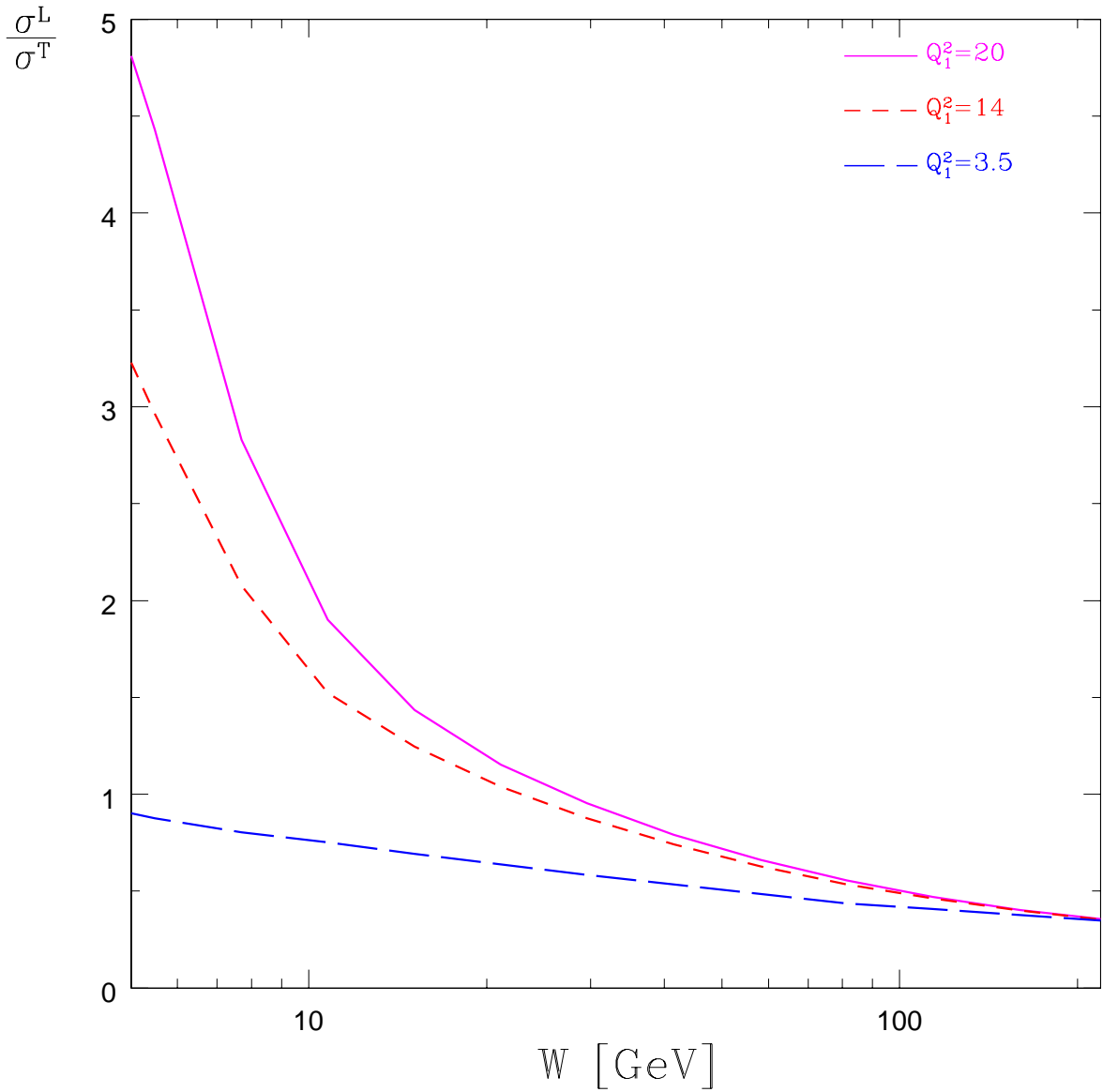,width=\textwidth}
  \caption[]{\parbox[t]{
             0.80\textwidth}{\small \it ${\displaystyle\frac{\sigma^L}{\sigma^T}}$ for the
             case $Q^2_2=0$.}}
\label{ltreal}
\end{center}
\end{figure}

\begin{figure}
\begin{center}
  \epsfig{file=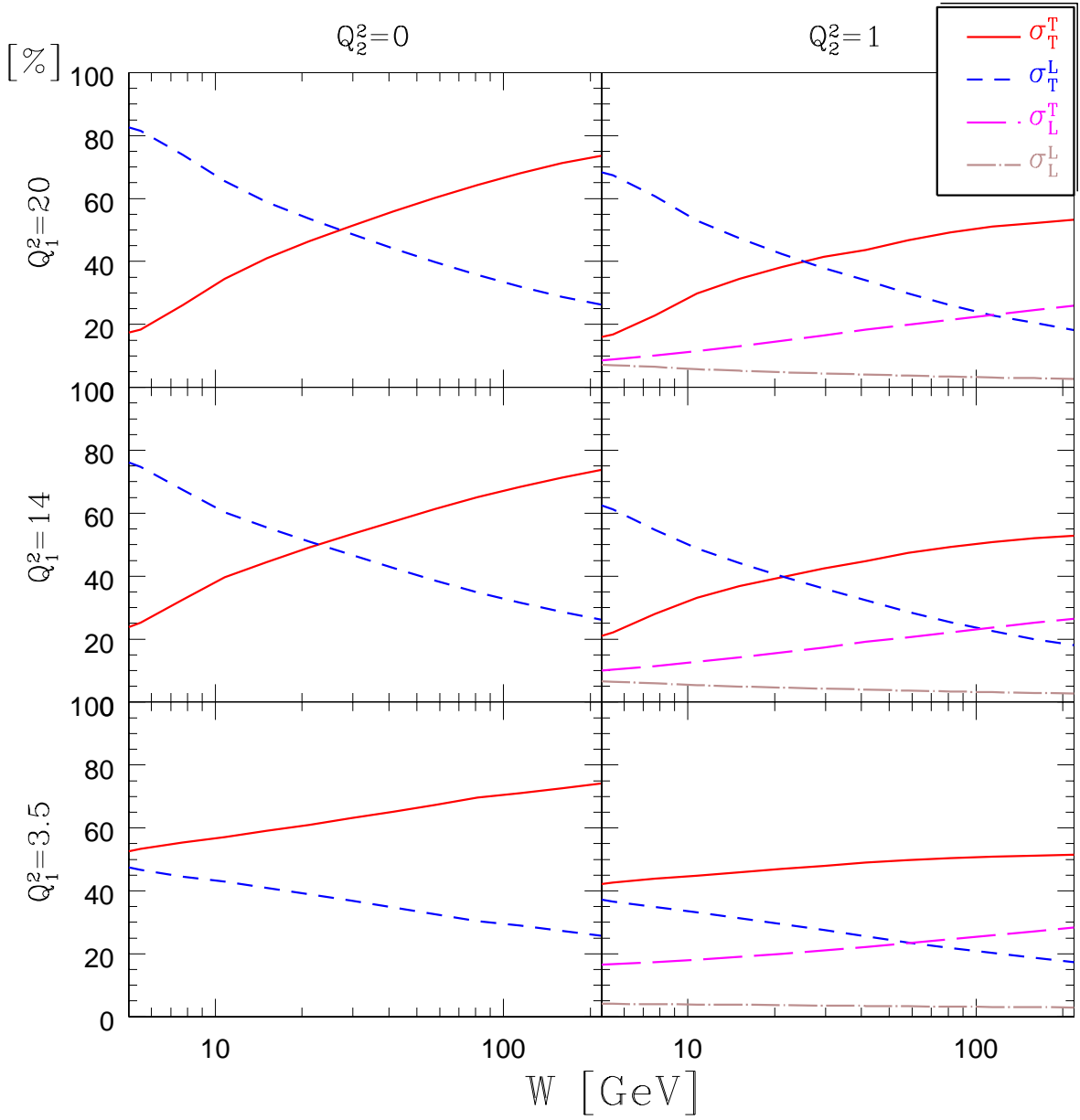,width=\textwidth}
  \caption[]{\parbox[t]{
             0.80\textwidth}{\small \it The relative contribution of
             $\sigtt,\,\sigtl,\,\siglt$ and $\sigll$ as a percentage from
             $\sigma(\gamma^*\gamma^*)$ for the case of
             $Q^2_2=0$ (left column) and $Q^2_2= 1\gevs$ (right column). The
             values of $Q^2_1$ are shown on the left hand side of the figure.}}
\label{ltvirt}
\end{center}
\end{figure}

\newpage
\newcommand{\refbrake}{\\\hspace*{2mm}}

\end{document}